\documentclass[10pt]{article}

\usepackage{amsmath}
\usepackage{amssymb}
\usepackage{graphicx}
\usepackage{cite}
\usepackage{color}
\usepackage{subfigure}
\usepackage{rotating}
\usepackage{amsmath}
\usepackage{amssymb}
\usepackage{amsfonts}
\usepackage{color}
\usepackage{float}

\definecolor{light-gray}{gray}{0.90}

\begin{document}

\begin{flushleft}

{\LARGE \textbf{Cancer therapeutic potential of combinatorial immuno- and vaso-modulatory interventions}}
\\
\vspace{5mm}

H. Hatzikirou$^{1,2,3,*,\dagger}$, J. C. L. Alfonso$^{1,2,*}$, S. M\"{u}hle$^{4}$, C. Stern$^{4}$, S. Weiss$^{4,5}$ and M. Meyer-Hermann$^{3,6}$
\\
\vspace{3mm}

$^{1}$ Center for Advancing Electronics, Technische Universit\"{a}t Dresden, 01062 Dresden, Germany.
\\
\vspace{1mm}

$^{2}$ Center for Information Services and High Performance Computing, Technische Universit\"{a}t Dresden, 01062 Dresden, Germany.
\\
\vspace{1mm}

$^{3}$ Department of Systems Immunology and Braunschweig Integrated Centre of Systems Biology, Helmholtz Center for Infectious Research, Inhoffenstr 7, 38124 Braunschweig, Germany.
\\
\vspace{1mm}

$^{4}$ Molecular Immunology, Helmholtz Center for Infectious Research, Inhoffenstr 7, 38124 Braunschweig, Germany.
\\
\vspace{1mm}

$^{5}$ Institute of Immunology, Medical School Hannover, Carl-Neuberg-Strasse 1, 30625 Hannover, Germany.
\\
\vspace{1mm}

$^{6}$ Institute for Biochemistry, Biotechnology and Bioinformatics, Technische Universit\"{a}t Braunschweig, 38106 Braunschweig, Germany.
\\
\vspace{3mm}

$^{*}$ These authors equally contributed to this work.\\
$^{\dagger}$ Corresponding author: haralambos.hatzikirou@tu-dresden.de
\vspace{3mm}

The authors declare that they have no competing interests.

\end{flushleft}


\section*{Abstract}

Currently, most of the basic mechanisms governing tumor-immune system interactions, in combination with modulations of tumor-associated vasculature, are far from being completely understood. Here, we propose a mathematical model of vascularized tumor growth, where the main novelty is the modeling of the interplay between functional tumor vasculature and effector cell recruitment dynamics. Parameters are calibrated on the basis of different {\it in vivo} immunocompromised Rag1$^{-/-}$ and wild-type (WT) BALB/c murine tumor growth experiments. The model analysis supports that tumor vasculature normalization can be a plausible and effective strategy to treat cancer when combined with appropriate immuno-stimulations. We find that improved levels of functional tumor vasculature, potentially mediated by normalization or stress alleviation strategies, can provide beneficial outcomes in terms of tumor burden reduction and growth control. Normalization of tumor blood vessels opens a therapeutic window of opportunity to augment the antitumor immune responses, as well as to reduce the intratumoral immunosuppression and induced-hypoxia due to vascular abnormalities. The potential success of normalizing tumor-associated vasculature closely depends on the effector cell recruitment dynamics and tumor sizes. Furthermore, an arbitrary increase of initial effector cell concentration does not necessarily imply a better tumor control. We evidence the existence of an optimal concentration range of effector cells for tumor shrinkage. Based on these findings, we suggest a theory-driven therapeutic proposal that optimally combines immuno- and vaso-modulatory interventions. \\


\noindent {\bf Keywords:} tumor-immune system interactions, murine tumor growth experiments, vascular normalization, immuno-modulatory interventions, mathematical modeling.

\section*{Major Findings}

We propose a tumor-effector cell recruitment model, where the main novelty is the low dimensional modeling of the complex interplay between functional tumor-associated vasculature and effector cell recruitment dynamics. Improved tumor vascularization, either via normalization or due to a microenvironmental stress alleviation strategy, is predicted to open a therapeutic window of opportunity in the treatment of cancer. Normalizing tumor vasculature as a potential therapeutic target closely depends on the immune recruitment dynamics and tumor sizes. Moreover, arbitrary low or high initial concentrations of effector cells can be detrimental to clinical outcomes, which evidences the existence of an optimal concentration range of effector cells that is crucial to obtain tumor control.


\section*{Introduction}

Suppression of tumor angiogenesis has been recognized for more than four decades as a therapeutic target to treat solid tumors, as well as a complementary method of cancer prevention \cite{Folkman1972, Jain2001, Miao2012}. Traditionally, antiangiogenesis strategies attempt to inhibit new blood vessels formation within the tumor microenvironment, while at the same time reducing as much as possible the functionality of existing tumor-associated vasculature. A wide array of drugs to inhibit tumor angiogenesis, such as bevacizumab (Avastin, Genentech/Roche), a ligand-trapping monoclonal antibody against VEGF, and the inhibitors of multiple protein kinase targets, sorafenib (Nexavar, Bayer) and sunitinib (Sutent, Pfizer), has been successfully used in the clinic to treat solid tumors \cite{Argyriou2009}. However, different preclinical and clinical trials reveal that angiogenesis inhibitors alone have limited efficacy and only offer a modest survival benefit in a reduced cancer patient population \cite{Ebos2011, Jayson2012}. The clinical benefits of antiangiogenesis treatments are not only limited in terms of tumor shrinkage, vasculature destruction and patient survival, but also are restricted by transient effects, insufficient efficacy or development of treatment resistance \cite{Jain2001, Jain2005, Jain2013}. There are increasing evidences that the antitumor activity of most antiangiogenic drugs only became clinically significant in combination with conventional therapeutic modalities such as radiotherapy, chemotherapy or immunotherapy \cite{Duda2007, Huang2012, Huang2013}.

Tumor-associated vasculature exhibits opposing effects on tumor growth, invasion and progression \cite{Jain2001, Jain2005, Jain2013}. On the one hand, blood vessels are required for delivery of oxygen and nutrients to the tumor, which promotes its growth, and they provide malignant cells with a way to spread, invade and metastasize into other healthy areas of the body \cite{Jain2005, Ferrara2005, Weis2011, Farnsworth2014}. On the other hand, blood vessels reoxygenate the tumor, which increases its radiosensitivity, and improve delivery of cytotoxic drugs to the tumor bulk as well as infiltration of immune cells into the tumor parenchyma \cite{Jain2005, Fridman2012, Huang2013, Junttila2013}. In view of these opposing effects, the right treatment of a cancer patient relies on a mechanistic understanding and quantitative analysis of the cancer-mediated processes involved. This demands for a mathematical model of tumor growth that considers the interplay between functional tumor vasculature and immune cell recruitment dynamics.

There are several mathematical models of tumor growth, where some forms of the immune system dynamics were included \cite{Bellomo2008, Arciero2004, Matzavinos2004, DOnofrio2005, RobertsonTessi2012}. Different immuno-oncology studies reveal that tumors can survive in a microscopic and undetectable dormant state \cite{DOnofrio2005, Caravagna2010, Matzavinos2004}, as inferred from clinical data \cite{Koebel2007}. This equilibrium can be disrupted by sudden events affecting the immune system. Indeed, the neoplasm develops several strategies to circumvent the action of the immune system \cite{Pardoll2003, Dunn2004}, which could result in tumor evasion and regrowth \cite{Dunn2004}. Sustained oscillations by the immune system have been observed both in its healthy state and pathological situations \cite{Stark2007, Coventry2009, Caravagna2010}. An elegant mathematical model of the dynamics of immunogenic tumors revealed that tumor-immune system interactions can produce cyclic behaviors \cite{Kuznetsov1994}. In fact, the presence of an immune component in mathematical models has been described to be crucial for reproducing observed phenomena such as tumor dormancy and oscillatory growth \cite{Kuznetsov1994, Pillis2006, Caravagna2010, DOnofrio2010}. This was complemented by a detailed discussion of the importance of including an immune component in tumor growth models, see \cite{Pillis2006} and references therein, and was subject to several reviews \cite{Araujo2004, Byrne2006, Roose2007, Bellomo2008, Eftimie2011, Wilkie2013}. Some of the models mainly focus on spatial aspects described either by partial differential equations or by cellular automatas or by realistic mechanical cell interactions \cite{Araujo2004, Schaller2005, Roose2007, Schaller2007}, while others focus on non-spatial dynamics described by ordinary differential equations \cite{Sachs2001, Eftimie2011}. Mathematical modeling provides a valuable theoretical framework to perform {\it in silico} experiments, as well as to verify assumptions and suggests modifications of existing theories. Despite years of research, and the vast amount of reported mathematical models, there are still many unanswered questions regarding the critical components of the tumor microenvironment and their role in the associated immuno-oncological mechanisms. However, models were rather successful in optimizing therapeutical protocols based on quantitative incorporation of experimental or patient data \cite{Pillis2006, Byrne2010, Jain2011, Corwin2013, Kempf2013, Alfonso2014a, Alfonso2014b, Jain2014}. The work presented here, follows this latter approach.

We intend to generate novel mechanistic insights not only on the interactions between the immune system and growing tumors, but also on the role of the functional tumor-associated vasculature. The main goal is to use the better understanding to suggest optimized treatment strategies. For that purpose, we combined a mathematical model of radially symmetric tumor growth \cite{Greenspan1976, Byrne1995, Cristini2003, Hatzikirou2012} with an immune cell model \cite{Kuznetsov1994} to propose a tumor-effector cell recruitment model. The main novelty is that the impact of functional tumor vasculature on the effector cell recruitment dynamics is included. Previous models of tumor-immune system interactions did not directly incorporate the effect of tumor vascularization, and thus underestimated the relevance of this crucial mechanism for tumor growth and treatment outcomes. Based on well-supported biological assumptions, we derive that tumor vasculature normalization opens a therapeutic window of opportunity for tumor growth control. The potential benefit of a vascular normalization strategy relies on a specific immune recruitment dynamic into the tumor microenvironment. We also find that an unlimited increase on the initial effector cell concentration does not necessarily imply a better tumor control. The proposed model is capable of providing a more complete view of vascularization effects on the immune responses to tumor growth, and it is therefore suited for analysis and prediction of treatment strategies.


\section*{Quick Guide to Model Equations and Assumptions}


\subsection*{A radial tumor growth model}

We consider a mathematical model of radial tumor growth which was originally proposed in \cite{Greenspan1976}, see also \cite{Cristini2003, Hatzikirou2012}. The tumor is assumed as an incompressible fluid flowing through a porous medium, where tissue elasticity is simplified. The tumor-host interface is considered to be sharp, and cell-to-cell adhesive forces are modelled as a surface tension at that interface. The tumor expands as a mass whose growth is governed by a balance between cell birth (mitosis) and death (apoptosis and necrosis). The mitotic rate within the tumor is assumed to be linearly-dependent on the nutrient concentration (oxygen, glucose, etc.) and is characterized by its maximal value $\lambda_M$ at the tumor-host interface. The death rate $\lambda_A$ is considered uniform within the tumor and assumed constant in time. The concentration of nutrients obeys a reaction-diffusion equation in the tumor volume, where nutrients are supplied from the functional tumor vasculature and consumed by the tumor cells at a uniform consumption rate (see the derivation of the mathematical model provided in the Supplementary Material for further details).

The resulting dimensional mathematical model for the tumor radius $R(t)$ dynamics, under the assumption of radial symmetry, is given by

\begin{equation}\label{eq:Radius_evolution}
    \frac{dR}{dt} = \frac{1}{3} (\lambda_M B - \lambda_A) R + \lambda_M (1-B) L_D \left( \frac{1}{\tanh (R/{L_D})} - \frac{L_D}{R} \right).
\end{equation}

This model given by Eq.~(\ref{eq:Radius_evolution}) can be interpreted as the temporal evolution of the average tumor radius, since radial symmetric growth is not a realistic behaviour. Notice that, multiplication of Eq.~(\ref{eq:Radius_evolution}) by $R^2$ describes the time evolution of the tumor volume (see the Supplementary Material for more details). The non-negative and dimensionless parameter $B$ represents the net effect of tumor-associated vasculature on the tumor radius dynamic. Moreover, $L_D$ is an intrinsic scale representing the average length of nutrient gradient, i.e.~supply, diffusion and consumption.


\subsection*{A tumor-effector cell recruitment model considering the impact of functional tumor-associated vasculature}

The proposed tumor-effector cell recruitment model is a combination of the radial tumor growth model given by Eq.~(\ref{eq:Radius_evolution}) and an effector cell model initially introduced in \cite{Kuznetsov1994}, see also \cite{DePillis2005, DOnofrio2005}. In addition, we consider the impact of functional tumor vasculature on the tumor growth and immune response dynamics. The system variables are the average tumor radius $R(t)$ and effector cell concentration $E(t)$ in the tumor vicinity. The model is formulated as a system of ordinary differential equations (ODEs) given by

\begin{eqnarray}
   \label{eq:TE_1} \frac{dR}{dt}&=&\frac{1}{3} (\lambda_M B - \lambda_A) R + \lambda_M (1 - B) L_D \left( \frac{1}{\tanh(R / L_{D})} - \frac{L_{D}}{R} \right) -  c E R f(R,\alpha),\\
   \label{eq:TE_2} \frac{dE}{dt}&=& r \frac{R^3}{K + R^3} E - \left( d_1 R^{3}  f(R,\alpha) + d_0 \right) E + \sigma,
\end{eqnarray}

\noindent where the time coordinate $t$ on the system variables has been omitted for notational simplicity. The non-dimensionalized version of the model is provided in the Supplementary Material. In words, the Eqs.~(\ref{eq:TE_1}) and (\ref{eq:TE_2}) can be respectively read as follows
\\
\\
\{change of tumor radius\} = \{net vascular tumor growth\} + \{avascular tumor growth\} - \{death of tumor cells due to effectors\}, \\
\\
\{change of effector cell concentration\} = \{immune recruitment\} - \{inactivation + death of effector cells\} + \{background rate of immune effector recruitment\}.
\\
\\
The expression $f(R,\alpha)=\frac{R^{\alpha - 1}}{R^{\alpha - 1} + 1}\in[0,1]$, for $\alpha \in [0,1)$, is a phenomenological scaling function that models the infiltration of effector cells into the tumor bulk through the existing functional tumor vasculature. This function modulates the terms related to tumor-effector cell interactions, such as killing of tumor cells due to effectors and inactivation of effectors due to their antitumor activity. As a particular case, we assume that $\alpha = B$, where $0 \leq B \leq 1$ represents the net effect of functional levels of the tumor-associated vasculature. In the limit of avascular tumor growth $B=0$, such tumor-effector cell interactions take place only at the tumor surface. At the other extreme, for $B=1$ effector cells can potentially interact with any cancer cell within the tumor bulk. Notice that such scaling functions have been also considered in the classical von Bertalanffy approach and more recently in allometric models \cite{Herman2011}. Alternatively, we can view the encountering of effector and tumor cells mediated by fractal vasculature as a percolation process within the tumor mass. In this context, the term $f(R,\alpha)$ scales the average effector-tumor cell encountering rate in a vascularized tumor. In particular, the scaling term should be proportional to the corresponding percolation cluster mass of system size $R$ and fractal dimension $3 \alpha$, where the latter depends on the structure of tumor-associated vasculature \cite{Stauffer1994}.


\subsection*{Model assumptions}

\begin{enumerate}
  \item The temporal evolution of the average tumor radius is considered, where invasive and diffusive tumor mechanisms are not taken into account.
  \item The death rate $\lambda_A$ of tumor cells reflects the lump effect of apoptotic and necrotic processes.
  \item Innate immunity or base immune surveillance is represented as a minimum presence of active effector cells at any time, even in the absence of cancer cells.
  \item Effector cells are recruited depending on the concentration of tumor cells according to Michaelis-Menten dynamics.
  \item The efficacy of immune killing depends on the ability of effector cells to penetrate the tumor bulk via the functional tumor-associated vasculature \cite{Huang2012, Huang2013}. With improved vascularization, the effectors kill tumor cells not only on the surface of the tumor, but also further inside.
  \item Effector cells die with a constant rate and get inactivated in dependence on their antitumor activity.
\end{enumerate}


\section*{Materials and Methods}


\subsection*{Experimental data of tumor growth in mice}

We now describe the murine experiments of {\it in vivo} tumor growth considered. BALB/c mice were purchased (Janvier, France). Rag1$^{-/-}$ mice were purchased from The Jackson Laboratory (C.129S7(B6)-Rag1$^{\mbox{tm1Mom}}$/J). Ig$\alpha^{-/-}$ mice were obtained from Michael Reth, Freiburg, Germany (Cd79a$^{\mbox{tm2Pld}}$). All recombinant mice were bred at the HZI and all experiments were performed with female, 8-12 weeks old mice if not stated differently and under approval of LAVES (Nieders\"{a}chsisches Landesamt f\"{u}r Verbraucherschutz und Lebensmittelsicherheit); Permission: 33.9-42502-04-12/0173. Unlike wild-type (WT) BALB/c mice, immunocompromised Rag1$^{-/-}$ mice cannot produce functional T- and B-cells, i.e. have no adaptive immune responses.

CT26 (ATCC CRL-2638) cells were cultured in IMDM supplemented with 10\% FCS, 100~U/ml penicillin and 100~$\mu$g/ml streptomycin, 50~$\mu$M 2-mercaptoethanol, 2~$\mu$M L-glutamine and maintained at 37~$^{\circ}C$ and 5\% $CO_2$. The F1A11 (H-2$^d$) cell line is a murine fibrosarcoma that expresses $\beta$-galactosidase ($\beta$-gal) and was obtained by transduction of spontaneously transformed BALB/c fibroblast cell line F1 with the LBSN retroviral vector \cite{Paglia1996}. Tumors CT26 and F1A11 were set by injecting $5 \times 10^5$~cells in 100~$\mu$l PBS subcutaneously (s.c.). Growth was monitored by caliper and the mean tumor radius evolution between days 4 to 9 was observed. Volume was calculated as $V = 4/3 \pi (h w^2) / 8$, where $h$ = height and $w$ = width.


\subsection*{Estimation of model parameters}

The experimental data corresponding to immunocompromised Rag1$^{-/-}$ and wild-type (WT) BALB/c mice are analyzed in two different phases. In the first phase (days 0 to 4) exponential avascular tumor growth is assumed \cite{Bru2004}. Using the tumor radius at those early times of growth, the net proliferation rate of cancer cells is estimated. The tumor radius evolutions in the second phase of growth (days 4 to 9) are then used to determine the remaining model parameter values. Because of uncertainties due to the finite number of observations, consisting in 15 and 20 individual experimental trials respectively, we have implemented a bootstrapping procedure for each dataset considered. The goal was to obtain reliable estimates of the summary statistics, where a resampling procedure has been performed. A set of bootstrap samples is artificially generated and represents the natural variations in the experimental datasets \cite{Davison1997}. More precisely, we have considered non-parametric bootstrap samples, i.e.~samples that have been generated from the original experimental datasets by drawing with replacements. The default size managed, also called cloud size, has been 100 000 bootstrap samples per dataset, where the resulting mean and standard deviation of such samples have been considered for parameter calibration. Another crucial factor to obtain precise model fitting is to consider different initial parameter values. Consequently, we started the fitting procedure from a large number of different random initial conditions, as well as by randomly perturbing the model parameter set to be estimated. Then, the solution with the lowest residual variance was selected, which thus provides the best fit.


\subsubsection*{Early exponential tumor growth phase (days 0 to 4)}

For a precise understanding of the early tumor growth phase predicted by our mathematical model, we work under the small tumor radius assumption and Taylor expand the r.h.s.~of Eq.~(\ref{eq:Radius_evolution}) accordingly. Keeping terms up to the first nonlinear term, we obtain the following approximation

\begin{equation}\label{eq:Radius_approximation}
    \frac{dR}{dt}\simeq\frac{1}{3}(\lambda_M -\lambda_A) R-\frac{\lambda_M (1-B)}{45 L_D^2}R^3.
\end{equation}

The very early tumor growth phase, dictated by the first term of this approximation, is always of exponential nature and does not depend on the parameter $B$, i.e.~the tumor vascularization effects, as reported also in \cite{Cristini2003, Bru2004, Hatzikirou2012}. Therefore, this initial phase depends exclusively on the net proliferation rate of cancer cells, $\lambda_p = (\lambda_M - \lambda_A)$. Considering the initial tumor radius in the immunocompromised Rag1$^{-/-}$ mice experiments of 0.40~mm, which corresponds to a tumor spheroid of $5 \times 10^5$~cells of about 10~$\mu$m in diameter, and the tumor radius at day 4 of 2.0~mm, see Fig.~\ref{fig:figure1}(A), we find that $\lambda_p \approx 1.20\pm0.1$~day$^{-1}$. This value is in agreement with previous estimates of growth rate (early stages) for murine CT26 spheroids \cite{Alessandri2013}. Later, the initial exponential tumor growth gets affected by tumor-associated vascularization, which starts playing a significant role when the two terms of the r.h.s. of Eq.~(\ref{eq:Radius_approximation}) have similar orders of magnitude. This occurs around a critical tumor radius $R_B$ derived from the condition

\begin{equation}\label{eq:R_c}
    \frac{1}{3}(\lambda_M -\lambda_A)R_B \approx \frac{\lambda_M (1-B)}{45 L_D^2}R_B^3.
\end{equation}

Since $(\lambda_M -\lambda_A) = \lambda_p > 0$ is necessary for the initial tumor growth, Eq.~(\ref{eq:R_c}) requires $0 \leq B < 1$, where $B = 0$ represents an avascular tumor and $B \rightarrow 1$ a fully vascularized tumor. This yields

\begin{equation}\label{eq:R_c_estimation}
R_B \approx L_D\sqrt{\frac{15 (\lambda_M -\lambda_A)}{\lambda_M (1-B)}}.
\end{equation}

Eq.~(\ref{eq:R_c_estimation}) provides an estimate of the tumor size at which nonlinearities, involving the vascularization mechanisms, start significantly influencing the initial exponential tumor growth. The effect of the nonlinear terms is to saturate the growth, i.e.~$dR/dt = 0$, when the tumor radius reaches $R_B$. Since only exponential growth is observed for small tumor radii, angiogenesis should be activated before the critical tumor radius is reached and suggests that $R_B$ is an upper bound for the initiation of angiogenic processes. The value of $R_B$ can be predicted provided the values of the parameters in Eq.~(\ref{eq:R_c_estimation}) are known. Assuming time-invariant values of the characteristic mitotic $\lambda_M$ and death $\lambda_A$ rates of cancer cells, and the relation $\lambda_p  \approx 1.20\pm0.1$~day$^{-1}$ together with the physiological plausible value $\lambda_M = 1/18 \,h \approx 1.34$~day$^{-1}$ for the mouse colon carcinoma cell line (or CT26 murine tumor cells) \cite{Alessandri2013, Delarue2013}, yields $\lambda_A \approx 0.14\pm0.1$~day$^{-1}$. The parameters $B$ and $L_D$, i.e. the functional tumor vasculature and intrinsic scale representing the average length of nutrient gradient, remain to be determined, which are estimated from the growth phase ranging between days 4 to 9.


\subsubsection*{Linear tumor growth phase (days 4 to 9)}

\begin{figure}[H]
\centering
 \includegraphics[width=0.85\textwidth]{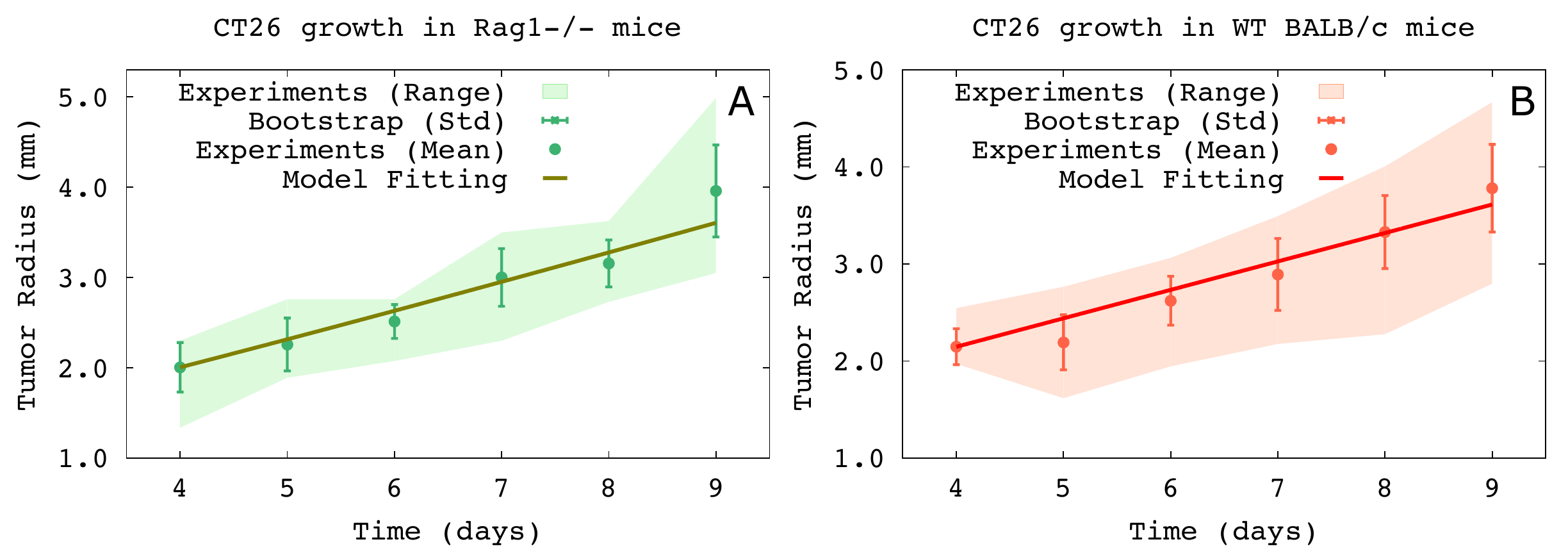}
  \caption{{\bf Linear fitting of the tumor radius evolution from murine experimental data.} (A) Tumor growth in immunocompromised Rag1$^{-/-}$ and (B) wild-type (WT) BALB/c mice at different time points are shown together with the linear fitting results of the model. The ranges of experimental data and standard deviations of the bootstrap samples are also shown.}
  \label{fig:figure1}
\end{figure}

From the experimental data corresponding to immunocompromised Rag1$^{-/-}$ and (WT) BALB/c mice at the stages of growth ranging between days 4 to 9, we observe an approximately linear evolution of the tumor radius, see Fig.~\ref{fig:figure1}. This observation is consistent with the long-term behavior of growth, where linear radial evolution is prominent as reported in \cite{Bru2004}. In Eqs.~(\ref{eq:TE_1})-(\ref{eq:TE_2}) this assumption is equivalent to assigning $B = \lambda_A/\lambda_M$. In particular, linear radial tumor growth in immunocompromised Rag1$^{-/-}$ mice requires $c = 0$ in Eq.~(\ref{eq:TE_1}), because Rag1$^{-/-}$ mice have no adaptive anticancer immune responses. Thus, the first and last terms on the r.h.s. of Eq.~(\ref{eq:TE_1}) vanish and allows to determine $L_D \approx 0.29 \pm 0.02$~mm. This value is consistent with the well-known characteristic nutrient diffusion length that ranges between 0.2~mm and 0.3~mm \cite{Frieboes2006, Cristini2008}. In Fig.~\ref{fig:figure1}(A) the linear fitting results for the Rag1$^{-/-}$ mice dataset are represented. The estimated parameter values allow for the calculation of $R_B \approx 1.12 \pm 0.08$~mm $> L_D$ from Eq.~(\ref{eq:R_c_estimation}) for a tumor in the early phase of growth which begins to be influenced by the tumor-associated vasculature. The following parameter values are taken from \cite{Kuznetsov1994, DOnofrio2005, DePillis2005, Su2009, DOnofrio2012}: $d_0 = 0.37$~day$^{-1}$, $d_1 = 0.01$~mm$^{-3}$~day$^{-1}$, $K = 2.72$~mm$^3$ and $\sigma = 0.13 \times 10^5$~cells~day$^{-1}$.

Radial tumor growth can be also assumed linear for the (WT) BALB/c mice dataset, see Fig.~\ref{fig:figure1}(B), which infers $dR/dt = V_{L} \approx 0.29 \pm 0.05$~mm/day constant in the time interval between days 4 to 9, and in agreement with growth velocity measured previously for murine CT26 tumors \cite{Seshadri2007}. Then, for $R \gg L_D$, Eq.~(\ref{eq:TE_1}) becomes

\begin{equation}\label{TE11} 
   \frac{1}{3} (\lambda_M B - \lambda_A) R + \lambda_M (1 - B) L_D - c E R^B = V_{L},
\end{equation}

\noindent where $\left( \frac{1}{\tanh(R / L_{D})} - \frac{L_{D}}{R} \right) \approx 1$, and for $0 \leq B \leq 1$ it follows that $\frac{R^B}{R^{(B-1)} + 1} \rightarrow R^B$ since $R^{(B-1)} \ll 1$. Considering that $R(t) = V_L (t - t_0) + R_0$, Eq.~(\ref{TE11}) provides the following evolution of the effector cell concentration $E(t)$ required for linear radial tumor growth

\begin{eqnarray}\label{eq:TE_11}
  E(t) = \frac{- V_L + \frac{1}{3} (\lambda_M B - \lambda_A) R(t) + \lambda_M (1 - B) L_D}{c R(t)^B}.
\end{eqnarray}

The above formula allows for the exact quantification of the initial concentration of effector cells $E_0$, as well as for the reduction of the number of free model parameters that enter in the fitting procedure, since the net proliferation $\lambda_p$ and death $\lambda_A$ rates of cancer cells, as well as $L_D$ have been estimated from the experimental data of immunocompromised Rag1$^{-/-}$ mice tumor growth. Accordingly, only three free model parameters are left in Eqs.~(\ref{eq:TE_1})-(\ref{eq:TE_2}), i.e. functional tumor vasculature $B$, immune cell recruitment rate $r$ and death rate of tumor cells due to effectors $c$, which are estimated from the (WT) BALB/c mice experiments imposing the expression (\ref{eq:TE_11}) during parameter calibration. Considering variations of the initial conditions, and randomly perturbing the parameter set to be estimated, the resulting parameter values are provided in Tab.~\ref{tab:params}. In Fig.~\ref{fig:figure1}(B) the linear fitting results for the experimental data of tumor growth in (WT) BALB/c mice are represented.

\begin{table}[H]
\centering
\begin{tabular}{lll}
\hline
Parameters & Values & Sources \\
\hline
  $c$ & 0.03 $\pm$ 0.002~cells$^{-1}$~day$^{-1}$ & {\it estimated} \\
  $r$ & 0.4 $\pm$ 0.01~day$^{-1}$ & {\it estimated} \\
  $K$ & $2.72$~mm$^3$ & \cite{Kuznetsov1994, DOnofrio2005, DePillis2005, DOnofrio2012} \\
  $d_1$ & 0.01~mm$^{-3}$~day$^{-1}$ & \cite{Kuznetsov1994, DOnofrio2005, DePillis2005, DOnofrio2012} \\
  $d_0$ & 0.37~day$^{-1}$ & \cite{Kuznetsov1994, DOnofrio2005, Su2009, DOnofrio2012} \\
  $\sigma$ & $0.13 \times 10^5$~cells~day$^{-1}$ & \cite{Kuznetsov1994, DOnofrio2005, DePillis2005, DOnofrio2012} \\
  $\lambda_M$ & 1.34~day$^{-1}$ & \cite{Alessandri2013, Delarue2013} \\
  $\lambda_A$ & 0.14 $\pm$ 0.1~day$^{-1}$ & {\it estimated} \\
  $L_D$ & 0.29 $\pm$ 0.02~mm & {\it estimated} \\
  $B$ & 0.18 $\pm$ 0.01 & {\it estimated} \\
  $E_0$ & $3.28 \pm 1.07 \times 10^5$~cells & {\it estimated} \\
\hline
\end{tabular}
\caption{
{\bf Model parameters.} Parameter values not found in the literature were estimated from {\it in vivo} experimental data of tumor growth in immunocompromised Rag1$^{-/-}$ and wild-type (WT) BALB/c mice.}
\label{tab:params}
\end{table}


\section*{Results}


\subsection*{The dual effect of functional tumor-associated vasculature}

The proposed mathematical model given by Eqs.~(\ref{eq:TE_1})-(\ref{eq:TE_2}) is used to investigate the effect of functional vasculature on the gross tumor growth. Tumor-associated vascularization enhances the supply of oxygen and nutrients, and therefore favours tumor growth and progression \cite{Ferrara2005, Farnsworth2014}. At the same time blood vessels facilitate the infiltration of effector cells into the tumor bulk, which potentially exterminate cancer cells \cite{Fridman2012, Junttila2013}. These opposing effects of tumor vascularization suggest the existence of an optimal tumor growth control, such that a quantitative modulation of functional vasculature can be considered as a suitable therapeutic target against cancer.

\begin{figure}[H]
\centering
 \includegraphics[width=0.75\textwidth]{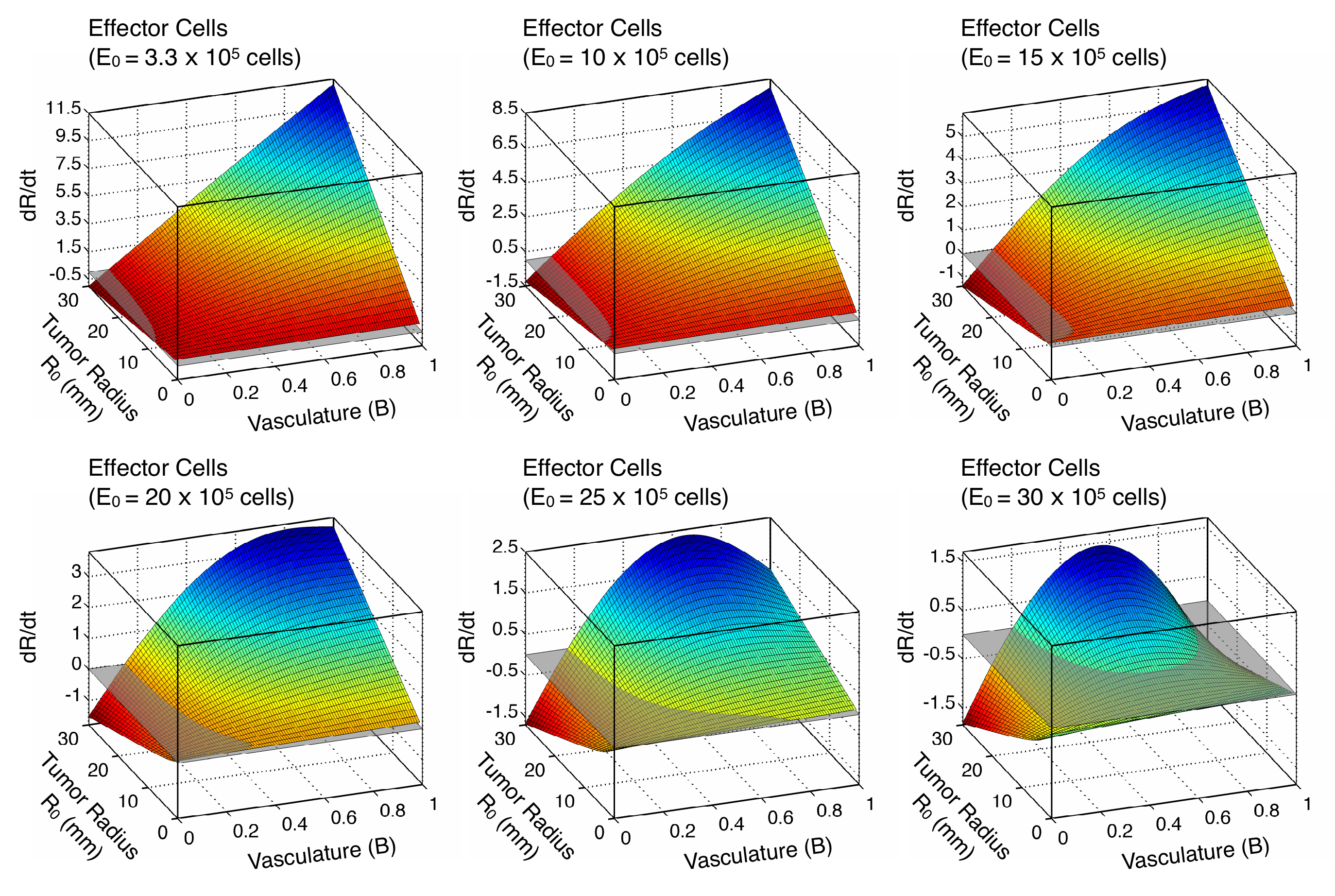}
  \caption{{\bf Short-term growth rate dependence on tumor size $R_0$ and functional vascularization $B$.} The radial tumor growth rate ($dR/dt$) is plotted against initial tumor radius $R_0$ and functional levels of tumor-associated vasculature $B$ for different initial concentrations of effector cells $E_0$. Remaining parameters in Eq.~(\ref{eq:TE_1}) are as in Tab.~\ref{tab:params}.}
  \label{fig:figure2}
\end{figure}

In order to gain insights into the impact of functional vasculature on the short-term tumor evolution, we vary the model parameter $B$ in Eq.~(\ref{eq:TE_1}). The radial growth rate $dR/dt$, which indicates whether the tumor radius grows (positive), decays (negative) or remains stable (zero), is monitored. Starting from different initial concentrations of effector cells $E_0$, Fig.~\ref{fig:figure2} shows the rate of tumor growth in dependence on the functional levels of vascularization $B$ and initial tumor sizes $R_0$. Above a specific threshold concentration of effector cells, a degree of vascularization exists that maximizes tumor growth. Moreover, above this threshold increasing vascular functionality reduces the tumor growth rate, since it allows a high number of effector cells to penetrate into the tumor parenchyma, see diagrams for $E_0 \geq 20 \times 10^5$~cells in Fig.~\ref{fig:figure2}. Tumor shrinkage $dR/dt < 0$ is possible for very low functional vascularization and large tumor radii, irrespective of the number of effector cells in the tumor vicinity. However, for small tumors and low initial concentrations of effector cells, reduction in the tumor burden cannot be obtained for deficient vascularization. Moreover, there exist critical $E_0$ values at which tumor shrinkage becomes also possible for well-vascularized tumors irrespective of their sizes, see diagram for $E_0 = 30 \times 10^5$~cells in Fig.~\ref{fig:figure2}.

\begin{figure}[H]
\centering
 \includegraphics[width=0.75\textwidth]{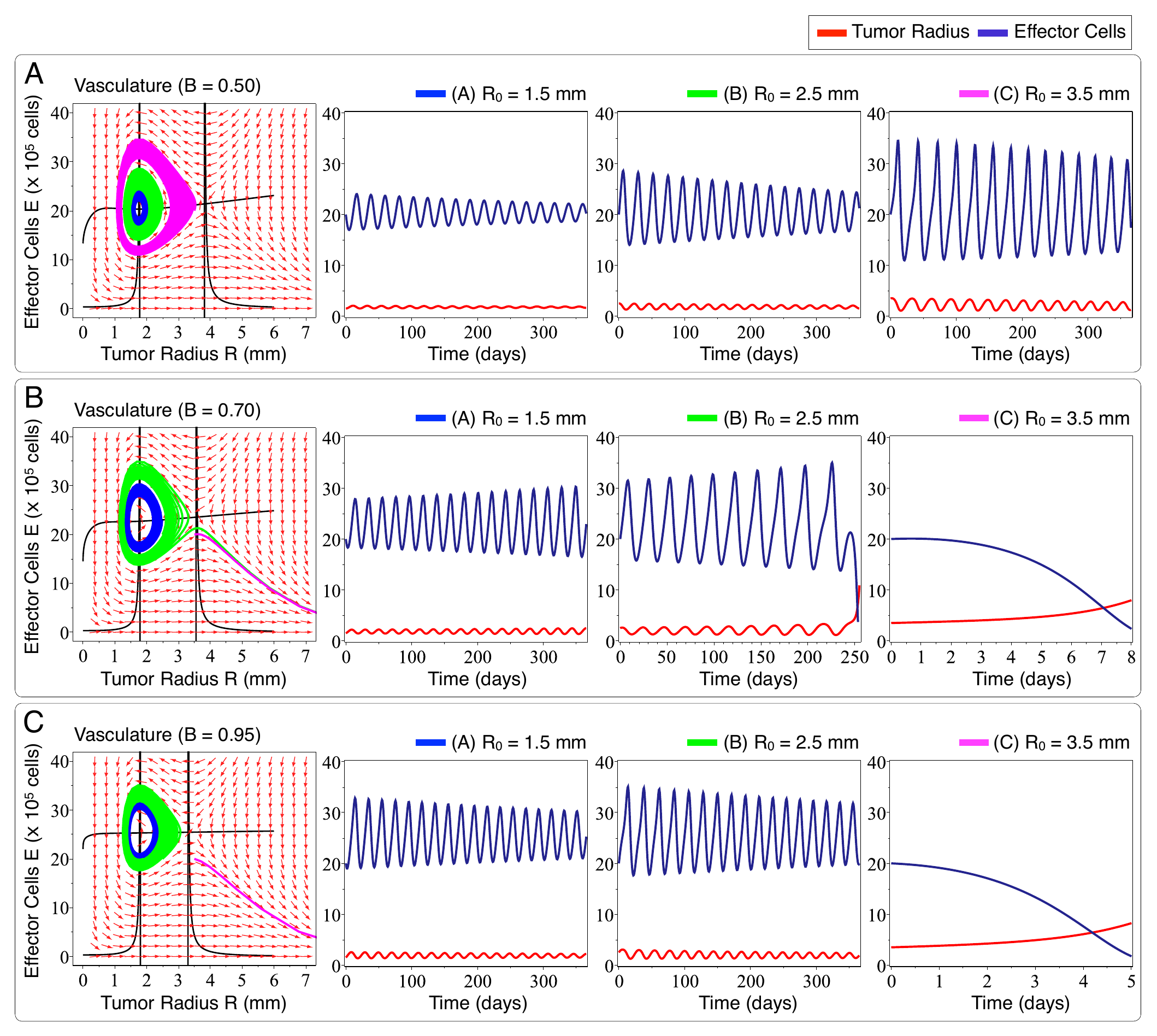}
  \caption{{\bf Tumor-immune phase portrait diagrams and long-term characteristic dynamics.} (Left) Phase portrait diagrams from top to bottom representing the tumor radius $R$ and concentration of effector cells $E$ time evolutions for different sets of initial conditions (colored curves) and levels of functional tumor vascularization $B$. The nullclines, i.e. zero-growth isoclines, of the dynamical system (black lines) are also represented. Each arrow provides information about the short-term (local) evolution of the corresponding initial conditions. (Right) Panels showing the corresponding time evolution of $R$ (red lines) and $E$ (blue lines) for $E_0 = 20 \times 10^5$~cells and different values of the initial tumor radius $R_0$ for each value of $B$ considered in the corresponding phase portrait diagrams on the left. The immune recruitment rate was $r = 0.59$~day$^{-1}$ and remaining parameters are as in Tab.~\ref{tab:params}.}
  \label{fig:figure3}
\end{figure}

A more systematic investigation of the dependence of long-term model dynamics given by Eqs.~(\ref{eq:TE_1})-(\ref{eq:TE_2}) on the set of initial conditions ($E_0$, $R_0$) is conducted through a tumor-immune phase portrait analysis, see Fig.~\ref{fig:figure3}. For functional vascular levels of $B = 0.50$, long-term tumor control is possible, in the sense of damped oscillations, for all values of $R_0$ considered, see Fig.~\ref{fig:figure3}(A). Tumor dormancy is described by decaying oscillations converging to a steady-state, that once obtained is maintained for a long period of time. However, at larger values of $R_0$ the amplitude of the oscillations in $E(t)$ and $R(t)$ increases and reaches uncontrolled growth above a threshold for $R_0$ which determines the tumor fate. For a higher tumor-associated vascularization, e.g. $B = 0.70$, long-term tumor control cannot be achieved for any set of values of the model parameters considered, see Fig.~\ref{fig:figure3}(B). In contrast, for well-vascularized tumors, e.g. $B = 0.95$, a window of opportunity for tumor control is predicted, see Fig.~\ref{fig:figure3}(C). As far as $R_0$ stays below a threshold value, tumor evolution can be controlled. At high and low vascularization levels, tumor eradication is not possible for sufficiently large tumors, revealing the relation between tumor sizes and the potential therapeutic benefit of targeting the tumor vascular network. The bifurcation analysis on model parameters $B$ and $r$, i.e. the functional tumor vasculature and immune recruitment rate, is provided in the Supplementary Material for further details.

\subsection*{Critical radius and control range of effector cell concentration determine tumor fate}

Fig.~\ref{fig:figure3} reveals information about the existence of different dynamical regimes of the tumor-effector cell recruitment system described by Eqs.~(\ref{eq:TE_1})-(\ref{eq:TE_2}) for the set of parameter values considered. Based on the analysis of the critical fixed points and their positions in the phase space (see the phase space and critical fixed points analysis provided in the Supplementary Material), we obtain that a spiral and a saddle point are present. At the spiral point the tumor radius oscillates around a small tumor size, i.e. a dormancy-associated equilibrium point. This implies a concentration range of effector cells that keep tumor under control for a period of time that depends on the point stability. Moreover, there exists a critical tumor radius $R_S$, which coincides with the saddle point, that for tumors of equal or larger sizes always grow uncontrollably and the immune system is completely suppressed. However, below this threshold tumors are always controlled, or at least for some period of time in a transient dormant state, provided the immune responses are in a specific range. The questions that arise are to (i) analytically evaluate the critical tumor size $R_S$ that for $R_0 \geq R_S$ implies uncontrollable growth and (ii) estimate the range of initial effector cell concentration $E_0$ that allows for long-term tumor control.

\subsubsection*{Critical radius for tumor dormancy}

These results evidence that the saddle point position in the phase space separates uncontrolled tumor growth and immune-induced dormant states, as well as delineates the potential and limits of therapeutic benefits of immuno- and vaso-modulatory interventions. For large well-vascularized tumors, we derive an analytical expression that approximates the critical tumor radius $R_S$ corresponding to an upper bound estimation of the saddle point (see the estimation of the critical radius for tumor dormancy provided in the Supplementary Material for further details), which is given by

\begin{equation}\label{eq:ApproxR}
    R_S = \sqrt[3]{ \frac{3 \sigma c}{d_1 (\lambda_M - \lambda_A)} - \frac{2 (d_0 - r)}{d_1}}.
\end{equation}

\begin{figure}[H]
\centering
  \includegraphics[width=0.75\textwidth]{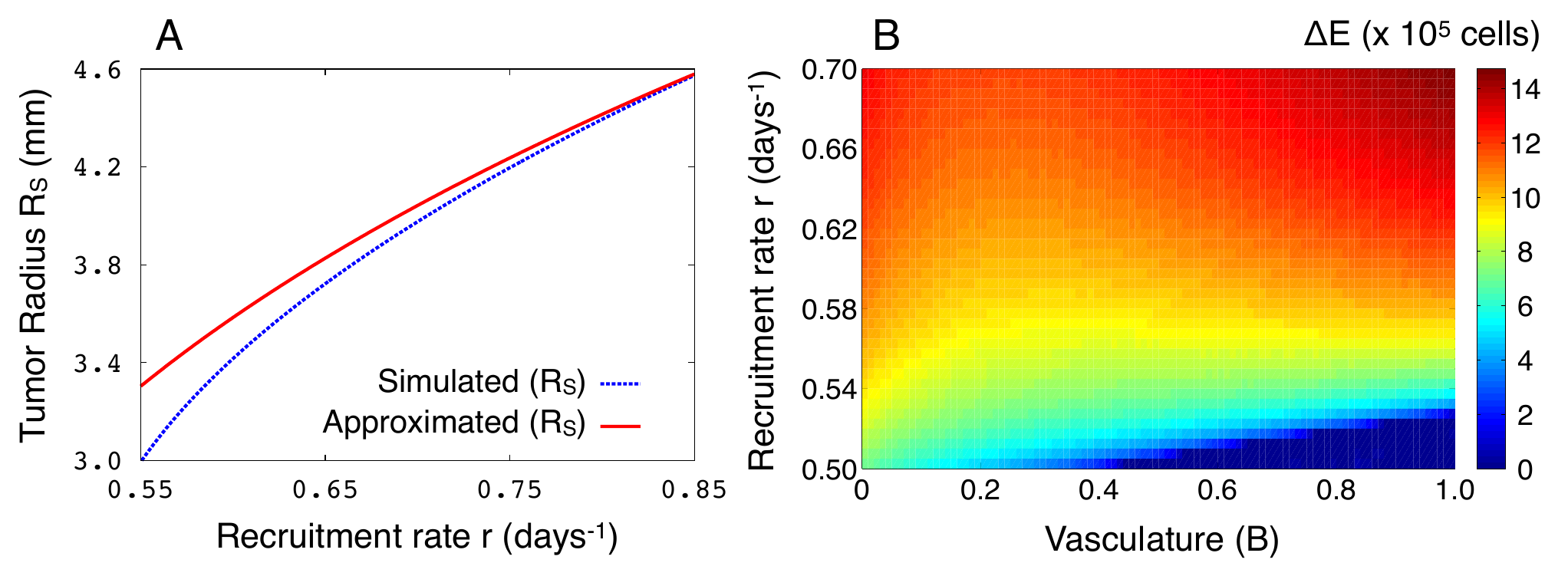}
  \caption{{\bf Dependence of the estimated critical tumor radius $R_S$, the tumor radius at the unstable saddle point and control range of effector cell concentration $\Delta E$ on the immune recruitment rate $r$.} (A) The tumor radius at the unstable saddle point in dependence on the parameter $r$ derived from model simulations (blue dashed line) and compared to the approximation of $R_S$ given by Eq.~(\ref{eq:ApproxR}) (red solid line) for a degree of functional tumor vascularization $B = 0.95$. (B) $\Delta E$ with respect to model parameters $r$ and $B$. Remaining parameters are as in Tab.~\ref{tab:params}.}
  \label{fig:figure4}
\end{figure}

We demonstrate that for values of the tumor radius equal or higher than $R_S$, the eigenvalues corresponding to the radial time evolution are positive. This implies that for radii beyond $R_S$, the tumor grows uncontrollably evading immune responses. The nonlinear system of equations (\ref{eq:TE_1})-(\ref{eq:TE_2}) was numerically solved using the Trust-Region Dogleg method \cite{Nocedal2006}. Fig.~\ref{fig:figure4}(A) shows the dependence of $R_S$ on the immune recruitment rate $r$ obtained by means of model simulations. $R_S$ represents an upper bound of the critical tumor size that determines the long-term tumor fate, and as $r$ increases the simulated and estimated values get closer. For $R < R_S$ long-term or at least transient tumor dormancy can be obtained, while $R \geq R_S$ results in tumor escape from the immune system surveillance. In the situation of tumors with low functional vascular levels, $R_S$ can be directly estimated by means of model simulations.

\subsubsection*{Control range of effector cell concentration regulates tumor evolution}

The efficacy of infiltrating effector cells to control the tumor growth is derived from the properties of the spiral point, which is always obtained for tumors with size lower than the critical threshold $R_S$. Tumor dormancy, and then neoplasia control, takes place when the tumor radius stays close to the spiral point for a long period of time. Therefore, the attraction range of the spiral point with respect to the E-axis provides useful information about the number of effector cells required for long-term tumor control. 

We claim that for arbitrary high or low initial concentrations of effector cells $E_0$, immune-induced tumor dormancy is not guaranteed. To illustrate the counterintuitive loss of tumor control for high or low initial numbers of effector cells, we focus at the phase portrait diagram of Fig.~\ref{fig:figure3}(B). There we observe that when the amplitude of the $E(t)$ oscillations exceeds a certain range, the tumor escapes from the immune system surveillance. Mathematically this is translated to the existence of a concentration range of effector cells $\Delta E$ that is crucial for the control of tumors with size smaller than the critical radius $R_S$. Details of $\Delta E$ calculation are provided in the Supplementary Material.

Fig.~\ref{fig:figure4}(B) shows the dependence of the control range of effector cell concentration $\Delta E$ on functional tumor-associated vasculature $B$ and immune recruitment rate $r$. In particular, we observe that $\Delta E$ increases for high enough immune recruitment rates and increasing tumor vascularization. However, at low values of $r \leq 0.54$~day$^{-1}$, $\Delta E$ becomes smaller as $B$ increases. The latter is intuitive considering that low immune recruitment rates could not provide enough effector cells to exploit the enhanced potential infiltration offered by increasing functional tumor vasculature. Therefore, the pro-tumoral effect of the functional vascular network overcomes its antitumoral action.


\subsection*{Therapeutic potential of immuno- and vaso-modulatory interventions}

Immune-mediated tumor dormancy depends not only on the effectiveness of the immune system responses, but also on the functional tumor-associated vasculature. In consequence, we investigate the impact of model parameters $B$ and $r$, i.e. the functional tumor vasculature and immune recruitment rate, on the overall therapeutic potential for long-term (5-years) model predictions. Therapeutic potential $TP(p): \Omega_p \rightarrow [0,1]$ is interpreted as the average tumor control, i.e.~tumor radius stays bounded, with respect to a certain regime of model parameter $p \in \{B,r\}$ given by

\begin{equation}\label{def:SP}
    TP(p) = \frac{\int\limits_{\Omega_p} H\big[R_S(p)-R(p,t^*)\big] dp}{\int\limits_{\Omega_p} H\big[R(p,t^*) \big] dp},
\end{equation}

\noindent where $t^*=1825$~days (5-years) is the target time, $R_S(p)$ is the tumor radius at the saddle point and $H(\cdot)$ is a Heaviside step function. The effect of the initial tumor radius $R_0$ and effector cell concentration $E_0$ have been also investigated. It should be noted that a 5-years tumor-free patient is typically considered in a complete remission stage, i.e.~permanent absence of disease, and thus cured.

\begin{figure}[H]
\centering
  \includegraphics[width=0.7\textwidth]{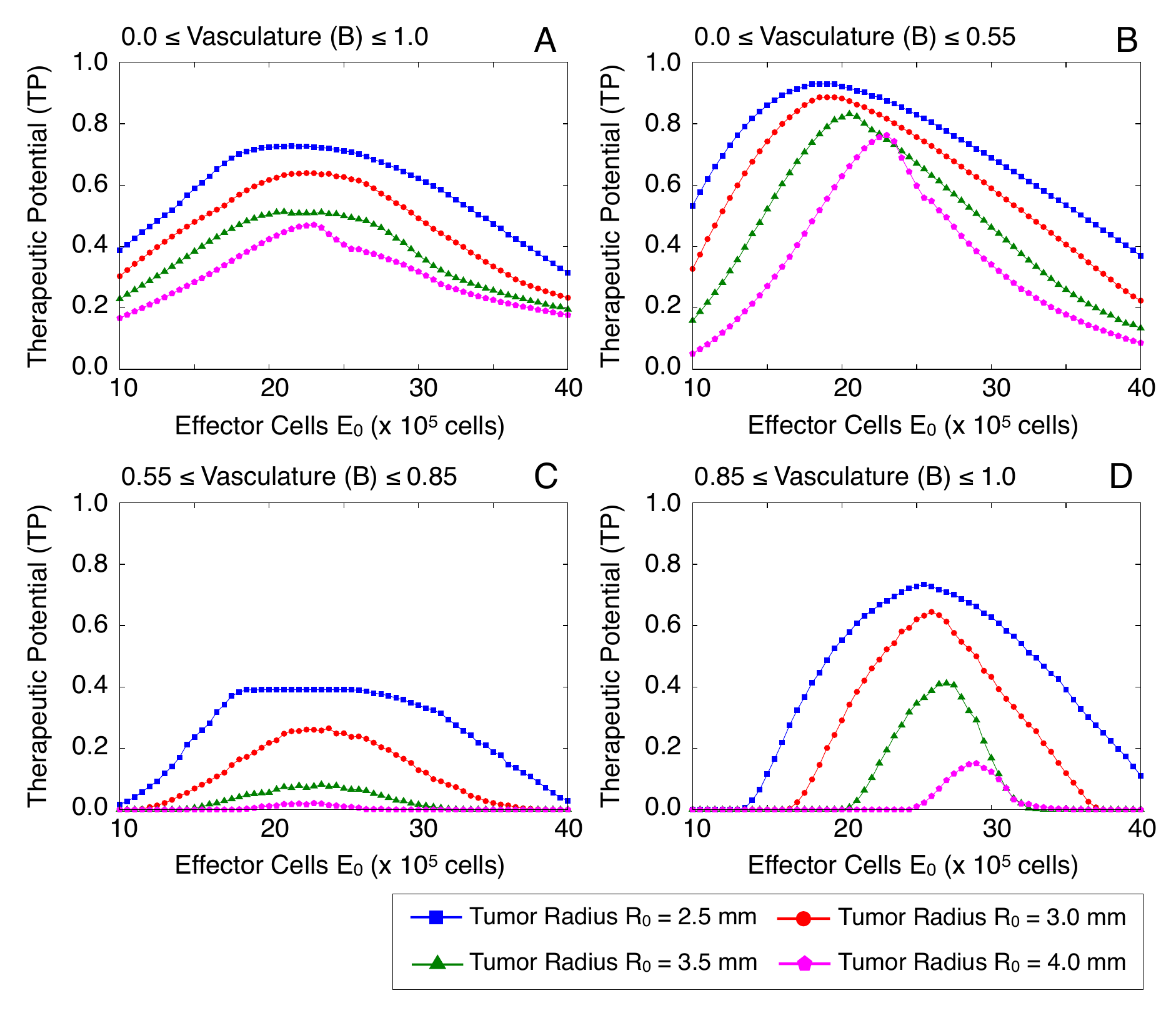}
  \caption{{\bf Therapeutic potential with respect to the initial concentration of effector cells $E_0$ and functional tumor vascularization $B$ for different values of the initial tumor radius $R_0$ and immune recruitment rate $r$}. The parameter $r$ varies between 0.50~day$^{-1}$ and 0.70~day$^{-1}$ for each value of $E_0$ considered. The y-axis represents the ratio of 5-years tumor control situations over all possible cases computed for the values of $r$ and $B$ in the ranges considered. Remaining parameters are as in Tab.~\ref{tab:params}.}
  \label{fig:figure5}
\end{figure}

Any increase in the initial tumor radius $R_0$ always reduces the therapeutic potential of the immune system to control tumor growth, see Fig.~\ref{fig:figure5}. However, an optimal concentration of effector cells for long-term tumor control is predicted irrespective of the degree of functional tumor vascularization $B$. This observation implies that an arbitrary increase in the initial concentration of effector cells does not necessarily result in a greater therapeutic benefit. It is counterintuitive that high values of $E_0$ do not enhance long-term tumor control. A large initial number of effector cells can induce a great initial tumor burden reduction that makes the tumor escape immune control by an undercritical immune stimulus. In mathematical terms, the tumor radius time evolution crosses the separatrix, escapes from the limit cycle and enters into the unlimited growth dynamic regime (see for instance the green curve in phase portrait diagram of Fig.~\ref{fig:figure3}(B), as well as the phase space and critical fixed points analysis provided in the Supplementary Material).

\begin{figure}[H]
\centering
  \includegraphics[width=0.7\textwidth]{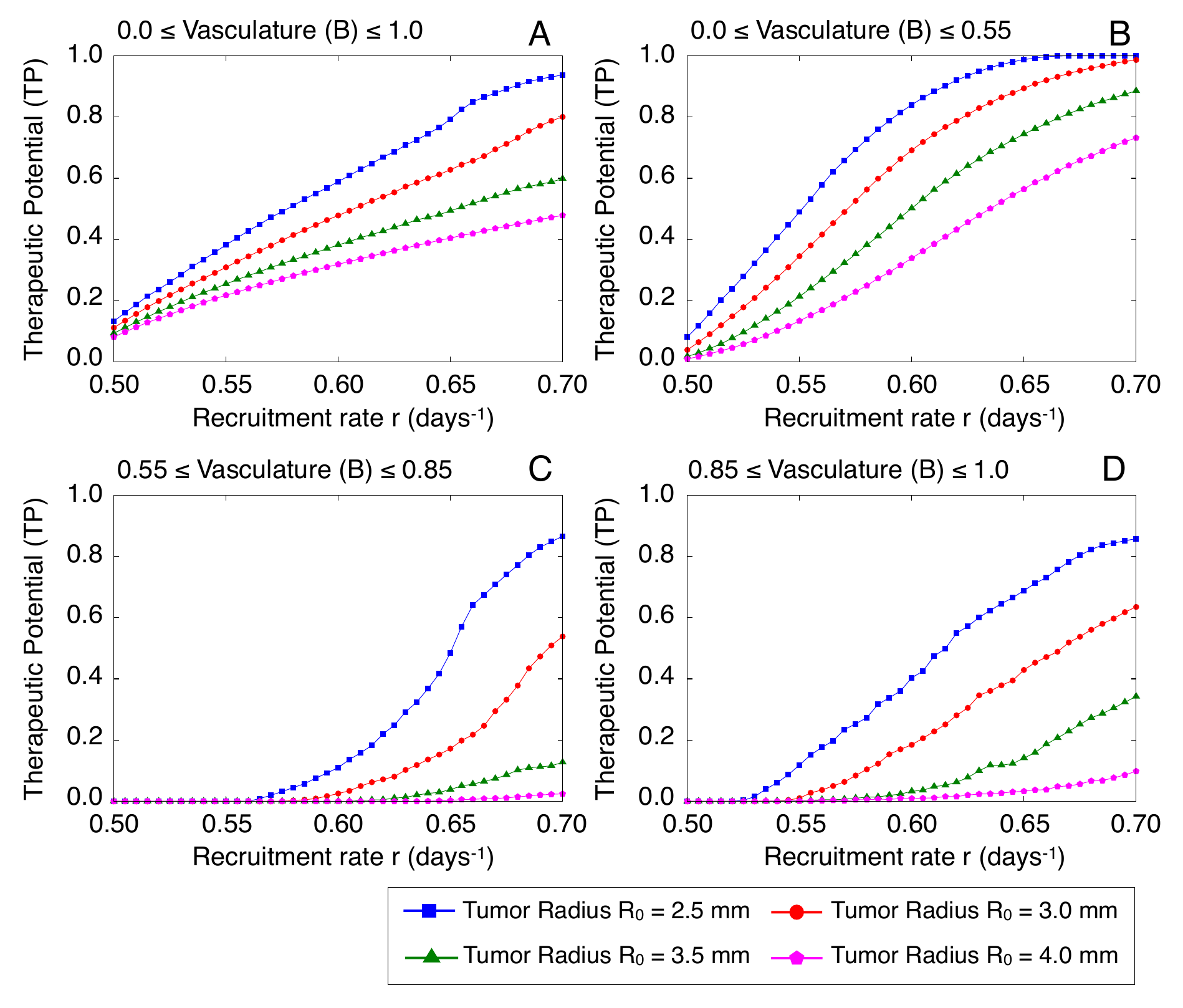}
  \caption{{\bf Therapeutic potential with respect to the immune recruitment rate $r$ and functional tumor vascularization $B$ for different values of the initial tumor radius $R_0$ and concentration of effector cells $E_0$}. The parameter $E_0$ varies between $10 \times 10^5$~cells and $30 \times 10^5$~cells for each value of $r$ considered. The y-axis represents the ratio of 5-years tumor control situations over all possible cases computed for the values of $E_0$ and $B$ in the ranges considered. Remaining parameters are as in Tab.~\ref{tab:params}.}
  \label{fig:figure6}
\end{figure}

In addition to the initial concentration of effector cells, the immune recruitment dynamics play an important role in long-term tumor control, see Fig.~\ref{fig:figure6}. Large values of the immune recruitment rate $r$ enhance the tumor control, irrespective of the tumor vascularization regime $B$. However, for high levels of functional tumor vasculature, i.e. $B > 0.55$, large tumors with $R_0 > 4$~mm are almost uncontrollable at any immune recruitment rate $r$, see Fig.~\ref{fig:figure6}(C,D). We also demonstrate that the initial concentration of effector cells required for long-term tumor control mostly depends on functional tumor vascularization (see the phase space and critical fixed points analysis provided in the Supplementary Material).

\subsection*{A theory-driven therapeutic proposal}

Model predictions provide new insights that can be used to suggest and design novel therapeutic strategies against tumor growth. The aim is to propose alternative or complementary ideas for less toxic cancer treatments compared to conventional therapeutic modalities, such as surgery, radiotherapy or chemotherapy. The latter cancer therapies are related to immunosuppressive effects that disallows the intention to exploit immune system responses.

An appropriate tumor profiling via medical imaging techniques and tissue biopsy is essential. Imaging allows for the estimation of the initial tumor size $R_0$ and its comparison with the critical radius $R_S$ that potentially determines the long-term tumor fate. For calculation of the precise $R_S$ value, the net proliferation rate of tumor cells could be extrapolated from biopsy samples by following the methodology proposed in \cite{Hyun2013}. Moreover, from the same pathology data the initial concentration of effector cells $E_0$ can be estimated, e.g. by counting the number of CD8$^+$ anti-tumor effector T cells. Then, the range of effector recruitment rate could be assumed comparable to the one estimated from our experimental data, see Tab.~\ref{tab:params}. For $R_0 < R_S$ mid- or long-term controlled tumor growth is possible by clinically influencing the interplay between immune recruitment dynamics and functional tumor-associated vasculature, as suggested by the therapeutic potential results in Figs.~\ref{fig:figure5} and \ref{fig:figure6}. In the situation of $R_0 \geq R_S$, tumor control could not be obtained by the findings obtained from the proposed model, and complementary treatment modalities are initially required.

Positive therapeutic responses, $dR/dt < 0$, can be achieved by increasing or decreasing the functional tumor vasculature. The latter is possible via antiangiogenic drugs, such as bevacizumab, sorafenib or sunitinib. However, low levels of tumor vascularization trigger hypoxic-induced phenomena, and this treatment choice typically fails after some short-term tumor burden reduction benefits \cite{Ebos2011, Jayson2012}. Hypoxic tumor regions promote either the secretion of alternative angiogenic factors, e.g. VEGF \cite{Hoeben2004}, or enhanced tumor cell migration \cite{Quail2013}, or suppressed immuno-activity allowing tumor cells to evade host immunosurveillance \cite{Huang2012, Huang2013}. Moreover, hypoxia is well known to reduce tumor cell sensitivity to radiation and chemotherapy \cite{Goel2011}. To avoid these adverse effects, and following our model predictions, an alternative strategy favouring high functional tumor vasculature could be adopted, i.e.~by increasing the model parameter $B$. This can be realized via normalizing the tumor vasculature \cite{Jain2001, Jain2005, Jain2013} or by means of a microenvironmental stress alleviation, i.e.~decompression of tumor vessels, see for instance \cite{Stylianopoulos2012}. Improved functional tumor vascularization allows for a wider range of effector cell types to infiltrate the tumor bulk and further exterminate cancer cells \cite{Fridman2012, Junttila2013}. This is supported by several preclinical studies demonstrating that vascular normalization increases infiltration of immune effector cells into tumors and substantially improves survival \cite{Manning2007, Hamzah2008, Shrimali2010}. Indeed, vascular normalization improves tumor blood vessel perfusion, reduces tumor hypoxia, enhances the delivery of cytotoxics and radiotherapy efficacy \cite{Huang2013}. Counterintuitively, normalization of tumor vasculature mitigates the metastatic risk since it is correlated to diminishing of epithelial-mesenchymal transition of cancer cells \cite{Agrawal2014}.

\begin{figure}[H]
\centering
  \includegraphics[width=0.65\textwidth]{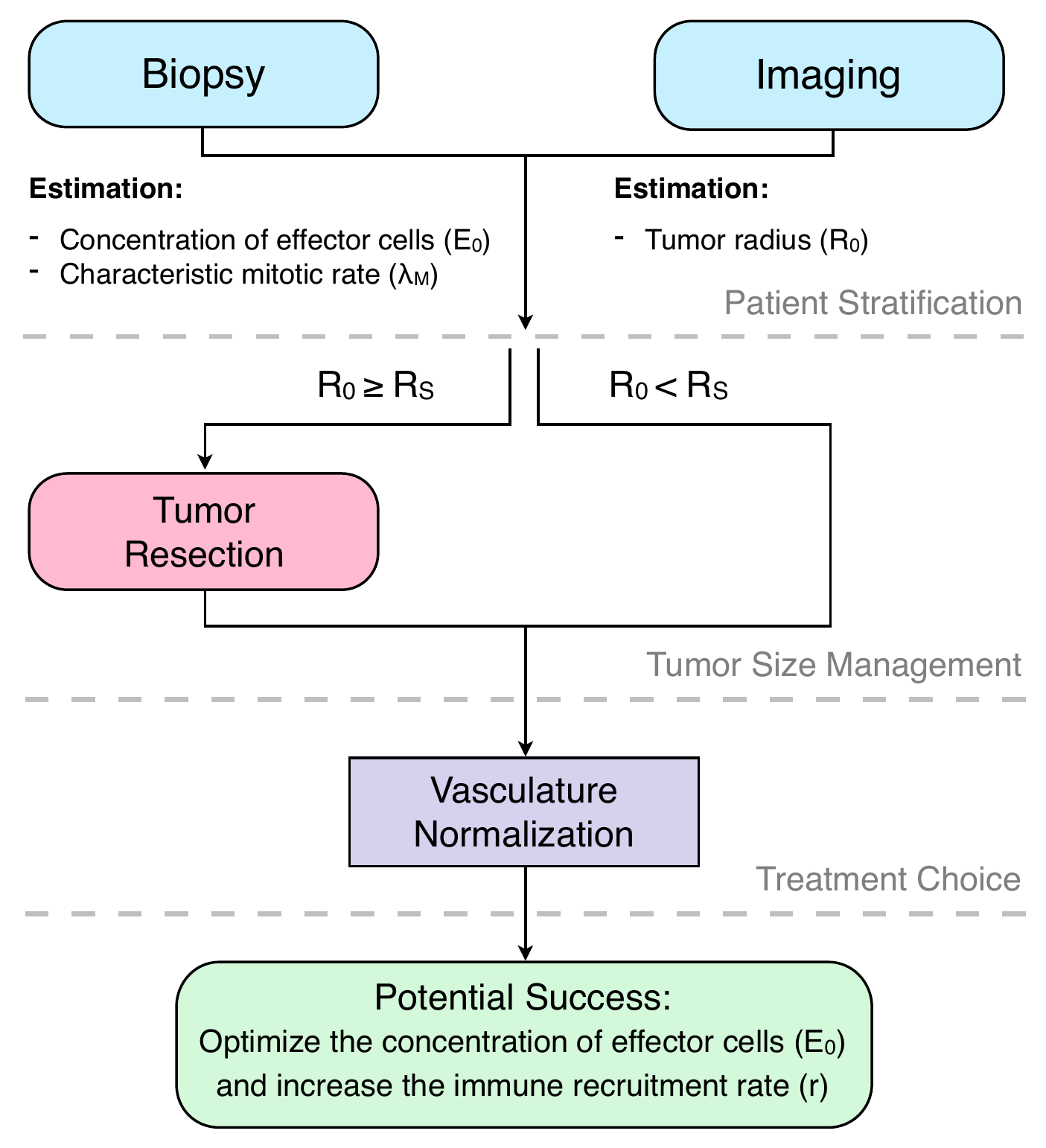}
  \caption{{\bf Theory-driven therapeutic proposal.} Medical imaging and biopsy allow for estimating the initial tumor radius $R_0$ and concentration of effector cells $E_0$, as well as the immune recruitment $r$ and tumor characteristic mitotic $\lambda_M$ rates. If the tumor is sufficiently small, $R < R_S$, a vasculature normalization strategy can be implemented to substantially improve tumor control. The potential benefit of normalizing the tumor vasculature is conducted through an increase of $r$ and optimizing $E_0$.}
  \label{fig:figure7}
\end{figure}

The model predictions shown in Fig.~\ref{fig:figure5} describe variations of the resulting therapeutic window of opportunity in terms of long-term tumor control for different functional levels of tumor-associated vasculature $B$, initial number of effector cells $E_0$ and tumor sizes $R_0$. A negative correlation between initial tumor sizes and mid- or long-term control is predicted. For sufficiently small tumors, optimal $E_0$ values range with higher therapeutic potentials exist. An increase of the immune recruitment rate results in a higher therapeutic potential and the window of opportunity for high functional levels of tumor vasculature still persists, see Fig.~\ref{fig:figure6}. Therefore, in addition to tumor vasculature normalization, applying a monitored immunotherapy protocol could further increase the expected therapeutic benefits. Fig.~\ref{fig:figure7} describes the complete theory-driven therapeutic proposal.

\section*{Conclusions}

In this work, we have explored the possibilities offered by mathematical modeling and computer simulations to gain insights into the interplay between functional tumor-associated vasculature and immune recruitment dynamics. We have combined a radially symmetric tumor growth and an effector cell model including the impact of vascularization on the immune system efficacy in killing tumor cells. The model parameter calibration was based on murine experiments of {\it in vivo} tumor growth. Model analysis revealed the existence of a tumor control window of opportunity when high functional levels of tumor vasculature are present. This can be biologically perceived since high vascularization allows for the penetration of effector cells into the tumor bulk and diminishes the adverse effects of hypoxia occurred due to low oxygenation. Improving tumor vascularization can be mediated by a normalization of blood vessel functionality \cite{Jain2001, Jain2005, Jain2013, Goel2011, Stylianopoulos2012}. Such an approach is successful only if the tumor size is below a critical threshold that depends on tumor cell metabolic needs. Furthermore, we predict that an arbitrary increase in the initial effector cell concentration does not necessarily imply a direct enhancement of tumor control, thus optimal values are needed. According to our model predictions when the initial reduction of tumor size exceeds a specific threshold, the tumor escapes immune control and relapses faster than the associated tumor-induced immune responses. Taking into account the dependency of tumor control on these various mechanisms, we have proposed an alternative less toxic therapeutic approach based on a personalized tumor profiling. This therapeutic proposal, based on normalizing the tumor vasculature, can be also potentially combined with chemotherapy and radiotherapy to enhance the therapeutic outcomes \cite{Jain2001, Jain2005, Goel2011}.

We conclude by pointing out a number of related future research directions, as well as the limitations of this work. The current results are based on murine experiments and therefore a parameterization on human data would be essential. Introducing tumor vascularization dynamics should be not only a natural, but also an insightful extension of the proposed mathematical model. Moreover, we need to further refine the modeling of the term $f(R,\alpha)$ and elaborate the dependence of the exponent $\alpha$ on the functional tumor vasculature $B$. Additionally, we could extend the notion of parameter $\alpha$ to reflect the dynamic feedback of effector cell activity on local tumor density fluctuations as a time-evolving fractal dimension. At this state, vascularization $B$ is considered as a static parameter in time. Making tumor vasculature dynamic would make sense to model further hypoxic effects, such as necrosis \cite{Macklin2007} or hypoxic-induced invasion \cite{Cristini2003}. Moreover, this will make possible to further investigate the dual effects of functional blood vessels on tumor growth dynamics when cytotoxic drugs are administered as part of chemotherapy. On the other hand, the immune system is much more complex than its description in the current model, involving a wide range of immune cell types and intricate interactions. In particular, the immune system is often regarded as acting to suppress tumor growth, however it is now clear that it can be both stimulatory and inhibitory \cite{Visser2006}. Recent advances have indicated that tumor-associated macrophages actively promote all aspects of tumor initiation, progression and development \cite{Hao2012}. Therefore, modeling the interplay between macrophages and tumor cells, while taking into account the functional tumor-associated vasculature and effector cell recruitment dynamics, may highlight new targets to develop novel anticancer strategies.


\section*{Author Contributions}

HH and JCLA conception and development of methodology; CS and SW conceived and designed the experiments; HH, JCLA, SM and MMH analysis and interpretation of data and model results; HH and JCLA administrative, technical and material support; SW and MMH study supervision; HH, JCLA, SW and MMH writing and revision of the manuscript. All authors read and approved the final manuscript.


\section*{Acknowledgments}
H. Hatzikirou and J. C. L. Alfonso gratefully acknowledge the support of the German Excellence Initiative via the Cluster of Excellence EXC 1056 Center for Advancing Electronics Dresden (cfAED) at the Technische Universit\"{a}t Dresden and the German Federal Ministry of Education and Research (BMBF) for the eMED project SYSIMIT (01ZX1308D). H. Hatzikirou and M. Meyer-Hermann were supported from the Human Frontier Science Program. M. Meyer-Hermann was supported by Measures for the Establishment of Systems Medicine projects SYSIMIT and SysStomach (01ZX1310C) by the Federal Ministry of Education and Research, Germany, and by iMed -- the Helmholtz Initiative on Personalized Medicine. C. Stern and S. Weiss gratefully acknowledge the support of the Deutsche Krebshilfe.


\newpage
\section*{Supplementary Material}


\subsection*{Derivation of the mathematical model of tumor growth}

We consider a non-necrotic tumor which volume growth results from a balance between cell mitosis and death, driven by the presence of nutrients, e.g. oxygen or glucose. In the absence of inhibitor chemical species, the spatio-temporal dynamic of the nutrient concentration $\sigma({\bf x},t)$ is modelled by the quasi-steady reaction-diffusion equation given by

\begin{equation*}
0= D\nabla^2 \sigma + \Gamma,
\end{equation*}

\noindent where $D$ is the diffusion coefficient and $\Gamma$ is the rate at which nutrients are supplied to the tumor volume $\Omega(t)$. The quasi-steady assumption is well supported by the observation that the diffusion time scale for oxygen or glucose is much lower ($\sim$ 1 minute) than the tumor cells doubling time ($\sim$ 1 day). The rate $\Gamma$ incorporates all sources and sinks in the tumor volume, and is based on the following phenomenological assumptions: \\

\noindent (1) Nutrients are homogeneously supplied by the vasculature at a rate $\Gamma_B = -\lambda_B \left( \sigma - \sigma_B \right)$, where $\sigma_B$ is the uniform nutrient distribution in the blood and $\lambda_B$ is uniform. \\

\noindent (2) Nutrients are consumed by tumor cells at a rate $\lambda\sigma$. This yields a rate $\Gamma$ given by

\begin{equation*}
\Gamma = -\lambda_B\left(\sigma-\sigma_B\right) - \lambda\sigma\,.
\end{equation*}

The tumor is modelled as an incompressible fluid which velocity field ${\bf u}$ in $\Omega$ satisfies the continuity equation given by

\begin{equation*}
\nabla\cdot{\bf u} = \lambda_P\,,
\end{equation*}

\noindent where $\lambda_P$ is the net proliferation rate of tumor cells that leads to volume growth (or shrinkage). This formulation rests on the additional assumptions described below: \\

\noindent (3) The tumor is modelled as a unique homogeneously distributed phenotype, meaning that all tumor cells behave in exactly the same manner. \\

\noindent (4) Tumor expansion exclusively depends on the net cell proliferation, and invasive processes, e.g. cell migration or diffusion, are not explicitly included. \\

\noindent (5) The model assumes that the density of tumor cells is constant and homogeneous within the tumor bulk. \\

\noindent (6) The net cell proliferation rate is assumed as follows

\begin{equation*}
\lambda_P = \lambda_M \frac{\sigma}{\sigma^\infty} - \lambda_A\,,
\end{equation*}

\noindent where $\sigma^\infty$ is the nutrient concentration outside the tumor volume, and the mitotic $\lambda_M$ and death rates $\lambda_A$ of tumor cells are uniform. \\

\noindent (7) The velocity is assumed to obey the Darcy's law, i.e. porous media flow, given by

\begin{equation*}
{\bf u} = - \mu \nabla P\,,
\end{equation*}

\noindent where $\mu$ is a (constant) cell motility parameter and $P({\bf x},t)$ is the pressure inside the tumor that is assumed to satisfy the Laplace-Young boundary condition at the interface, which corresponds to the following assumption: \\

\noindent (8) Cell-to-cell adhesive forces are modelled by a surface tension $\gamma$ at the tumor boundary. By introducing the intrinsic length scale $L_D = D^{1/2}/\left(\lambda_B + \lambda \right)^{1/2}$, we obtain an intrinsic relaxation time scale $\lambda_R^{-1} =(\mu\gamma)^{-1} L_D^3$. We consider these length and time scales to non-dimensionalize the mathematical model that can be rewritten, using bar-notation for dimensionless quantities, in terms of the modified nutrient concentration ${\bar \sigma}$ and pressure ${\bar p}$ defined by

\begin{equation*}
\sigma =\sigma^{\infty}\big(1-(1-B)(1-{\bar \sigma})\big)
\end{equation*}

\noindent and

\begin{equation*}
P =\displaystyle\frac{\gamma}{L_D}\left({\bar p}+\frac{1}{\lambda_R}\left(\lambda_M(1-B)(1-{\bar \sigma}) + (\lambda_A-\lambda_MB)\frac{\mathbf{\bar x} \cdot \mathbf{\bar x}}{2d}\right)\right),
\end{equation*}

\noindent where the parameter

\begin{equation*}
B = \frac{\sigma_B}{\sigma^\infty}\frac{\lambda_B}{\lambda_B+\lambda}
\end{equation*}

\noindent represents the net effect of vascularization.

Then, by using algebraic manipulations, the original dimensional problem can be reformulated in terms of two non-dimensional decoupled problems as follows

\begin{equation*}\label{eq_nutrient}
\begin{array}{l}
\nabla^2 \bar{\sigma} - \bar{\sigma} = 0,\\ \\
(\bar{\sigma})_{\Sigma} = 1;
\end{array}
\end{equation*}

\noindent and

\begin{equation*}\label{eq_pressure}
\begin{array}{l}
\nabla^2 \bar{p} = 0,\\ \\
(\bar{p})_{\Sigma} = \kappa  - \displaystyle\frac{1}{\lambda_R}(\lambda_A-\lambda_MB)\frac{(\mathbf{\bar x}\cdot \mathbf{\bar x})_{\Sigma}}{2d},
\end{array}
\end{equation*}

\noindent in a $d$-dimensional tumor separated from the host tissue by the interface $\Sigma$ of local curvature $\kappa$ that evolves with the normal velocity ${\bar v}=\mathbf{n}\cdot (\mathbf{\bar u})_{\Sigma}$, and $\mathbf{n}$ being the outward normal to $\Sigma$.

When considering evolution of a three-dimensional tumor that remains radially symmetric, problems above have analytical solutions that lead to the evolution equation for the dimensionless tumor radius ${\bar R}$ given by

\begin{equation*}
\frac{d{\bar R}}{d{\bar t}} = {\bar v} = \frac{1}{\lambda_R} \left(
\frac{1}{3}(\lambda_M B-\lambda_A){\bar R} + \lambda_M (1-B) \Big(\frac{1}{\tanh ({\bar R})}-\frac{1}{{\bar R}}\Big)
\right).
\end{equation*}

In particular, we have considered the dimensional version of the previous equation for the tumor radius $R = L_D {\bar R}$ evolving with respect to time $t = {\bar t} / \lambda_R$.


\subsection*{Tumor volume evolution}

It is instructive to compare the proposed mathematical model of tumor growth with other modelling approaches. In particular, we compare it with the well-known logistic growth model which has been extensively used in previous studies. To do that, we multiply both sides of Eq.~(1) in the manuscript by $3R^2$, and for large radii, i.e.~$R\gg L_D$, we obtain the following equation

\begin{equation*}
3R^2\frac{dR}{dt}=(\lambda_M B-\lambda_A)R^3+3\lambda_M (1-B) L_D R^2,
\end{equation*}

\noindent where we have that

$$\lim_{R\rightarrow \infty}\Big(\frac{1}{\tanh (R/{L_D})}-\frac{L_D}{R}\Big)=1.$$

Now, assuming that the tumor mass is $M = \rho V \propto R^3$ and $\rho$ is the tumor density per volume element, then we obtain an equation for the tumor mass given by

\begin{equation*}
\frac{dM}{dt}=(\lambda_M B-\lambda_A)M + 3\lambda_M (1-B) L_D M^{2/3},
\end{equation*}

\noindent where the term $M^{2/3}$ denotes the mass on the surface of volume $V$. In particular, for $B = 0$, i.e.~an avascular tumor, the above equation becomes similar to the well-known von Bertallanfy equation

\begin{equation*}
\frac{dM}{dt}=3\lambda_M  L_D M^{2/3}-\lambda_A M,
\end{equation*}

\noindent which represents the growth of a body that is supplied with nutrients only from its surface. The last equation is a kind of generalized logistic growth. In order to derive the classical logistic growth equation we need to restrict ourselves to a two-dimensional system. As a matter of fact, the main problems with the logistic model are:

\begin{enumerate}
\item It is valid only for very large tumors, but it fails to capture the dynamics of small tumors.
\item It is true only for fixed nutrient consumption within the tumor volume, i.e. no nutrient diffusion.
\item It is valid only for two-dimensional tumor evolution.
\end{enumerate}

Finally under the same assumption of $R\gg L_D$ and for a fixed concentration of effector cells $E_0$, we have that

\begin{equation*}
\frac{dM}{dt}=(\lambda_M B-\lambda_A)M + 3\lambda_M (1-B) L_D M^{2/3}-cE_0 M^{\frac{2+B}{3}}.
\end{equation*}

The last term represents the ``depth" of tumor-effector cell interactions with respect to the tumor-associated vasculature. Obviously, for the avascular limit $B=0$ such term indicated only surface interactions, while for $B=1$ effectors can kill cancer cells throughout the tumor bulk. 


\subsection*{Model non-dimensionalization}

The process of non-dimensionalization allows for (i) scaling the system of variables according to their characteristic quantities, (ii) writing the system by reducing as much as possible the number of model parameters, and (iii) identifying their appropriate units. To do that, we assume as reference length size the model parameter $L_D$, and the initial number of effector cells as reference concentration $E_{ref}$, for the variables $R$ and $E$ respectively. The diffusion length $L_D$ ranges in the interval [0.2~mm, 0.3~mm] \cite{Acker1984, Frieboes2006, Cristini2008}, and the effector cell characteristic concentration $E_{ref}$ is at the order of magnitude $10^5$~cells. The latter estimate is justified since the characteristic length scale of the system is at the order of 1~mm, and given that cells are commonly assumed with a diameter between 10~$\mu$m and 20~$\mu$m \cite{Delarue2013}, then for a volume of 1~mm$^3$ the concentration is at the order of magnitude $10^5$~cells. Moreover, the time scale is proportional to tumor cell deactivation, i.e.~$t' = (c' E_{ref}) t$, where the parameter $c'$ is provided below. 

Thus, the non-dimensional system of ordinary differential equations for the new variables $R' = R / L_D$ and $E' = E / E_{ref}$ is given by

\begin{eqnarray*}
   \label{eq:ndTE_1} \frac{dR'}{dt'}& = & \frac{1}{3} (\lambda_M' B - \lambda_A') R' + \lambda_M' (1 - B) \left(\frac{1}{\tanh(R')} - \frac{1}{R'} \right) - c' E' \frac{R'^B}{R'^{(B - 1)} + L_D^{(1-B)}}, \\
   \label{eq:ndTE_2} \frac{dE'}{dt'}& = & r' \frac{R'^3}{K' + R'^3} E' - \left( d_1' \frac{R'^3}{R'^{(1 - B)} + L_D^{(B-1)}} + d_0' \right) E' + \sigma',
\end{eqnarray*}

\noindent where the non-dimensional model parameters are summarized as follows

\begin{table}[H]
\centering
\begin{tabular}{ccccc}
  $c' = c ,$                       &   $\lambda_M' = \frac{\lambda_M}{c' E_{ref} },$   &   $\lambda_A' = \frac{\lambda_A}{c' E_{ref} },$   &                & \\
  \\
  $r' = \frac{r}{c' E_{ref} },$   &   $K' = \frac{K}{L_D^3},$                                 &   $d_1' = \frac{d_1 L_D^{(2+B)}}{c' E_{ref} },$     &  $d_0' = \frac{d_0}{c' E_{ref} },$   &   $\sigma' = \frac{\sigma}{c' E_{ref} ^2}.$
 \end{tabular}
\end{table}

It should be noted that the inactivation rate of effectors by encounters with tumor cells interestingly depends on the parameter $B$, i.e. the functional tumor vasculature, by means of the dimensionless parameter $d_1'$.


\subsection*{Phase space and critical fixed points}

\begin{figure}[H]
\centering
  \includegraphics[width=0.75\textwidth]{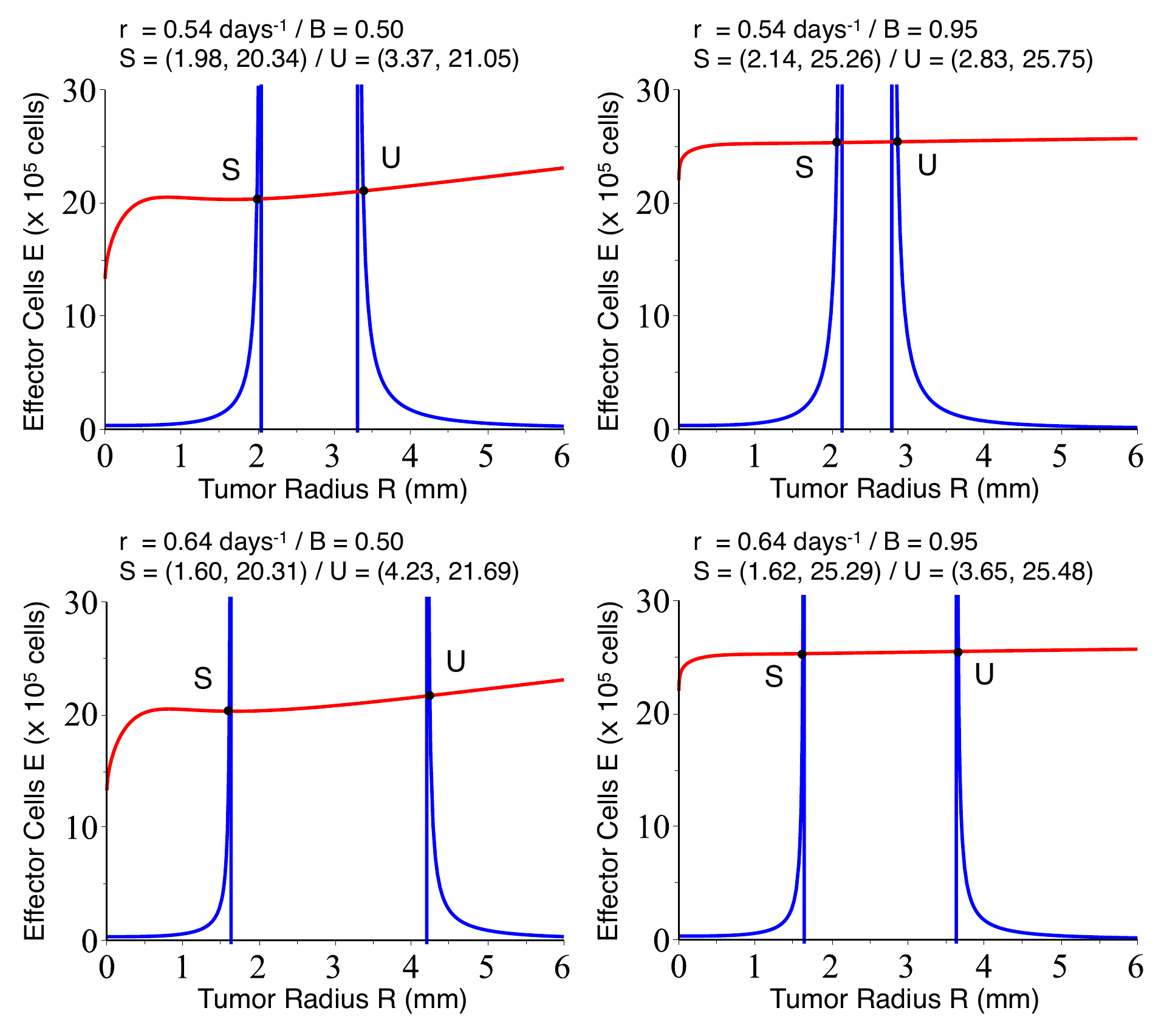}
  \caption{{\bf Nullclines and critical fixed points.} The nullclines $dE/dt=0$ (blue lines) and $dR/dt=0$ (red lines) are plotted for different values of model parameters $r$ and $B$, i.e. immune recruitment rate and functional tumor-associated vasculature respectively. The intersections of the nullclines represent the spiral and saddle points of the system, denoted by S and U respectively. Remaining parameters are as in Tab.~1 of the manuscript.}
  \label{fig:figure8}
\end{figure}

The nullclines, i.e. zero-growth isoclines, as well as critical fixed points corresponding to different model parameter values of the tumor-effector cell recruitment dynamical system ($R$-$E$ axis), are shown in Fig.~\ref{fig:figure8}. The intersections of the nullclines, i.e.~the curves along which $dR/dt = 0$ and $dE/dt = 0$, represent the critical fixed points of the system. Fig.~\ref{fig:figure8} shows that a spiral and a saddle point are present, which are represented by S and U on each phase diagram respectively. At the spiral point the time evolution of $R$ oscillates around a small tumor size, and the separatrix defines sharp boundary between tumor growth and shrinkage depending on the initial conditions, see also Fig.~3 in the manuscript. There exists a critical tumor size that coincides with the saddle point and above which tumors always grow uncontrollably and the immune system is completely suppressed. Below this threshold, tumors are always controlled when the spiral point is stable, or at least for some period of time if it is unstable, provided the immune responses are in a specific range. Therefore, the mathematical model is capable of explaining both tumor dormancy and escape from immunosurveillance.

\begin{figure}[H]
\centering
  \includegraphics[width=0.75\textwidth]{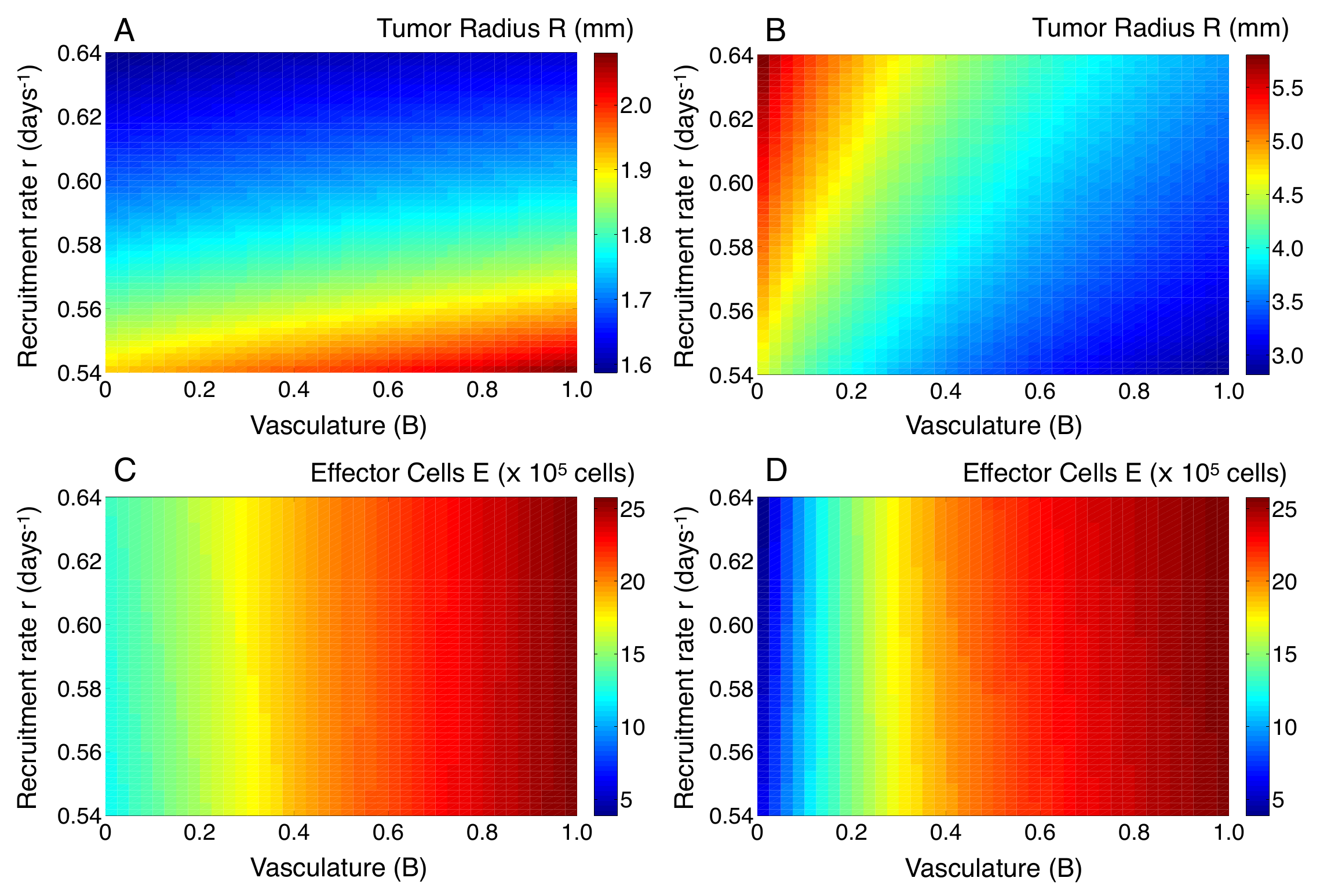}
  \caption{{\bf Position of the critical fixed points in the phase space}. Position dependence of the spiral and saddle points on different values of model parameters $r$ and $B$, i.e. immune recruitment rate and functional tumor-associated vasculature respectively. (A,B) Tumor radius and (C,D) effector cell concentration corresponding to the spiral (A,C) and saddle points (B,D). Remaining parameters are as in Tab.~1 of the manuscript.}
  \label{fig:figure9}
\end{figure}

A quick glance at Fig.~\ref{fig:figure8} reveals that the distance between the critical fixed points depends on the model parameter values. Fig.~\ref{fig:figure9} represents the position of the critical fixed points on the $R$-$E$ axis for different values of functional tumor vascularization $B$ and immune recruitment rate $r$. The variations of the tumor radius corresponding to the spiral point are always smaller, ranging between 1.59~mm to 2.08~mm, than for the unstable saddle point, varying from 2.82~mm to 5.80~mm, see Fig.~\ref{fig:figure9}(A,B) respectively. For $B \rightarrow 1$ and small values of $r$, the spiral and saddle points approach each other in the $R$-axis. Accordingly, this means that as the immune recruitment rate decreases and functional tumor vascularization increases, the tumor becomes less controllable, i.e.~the region for favorable initial conditions in terms of long-term tumor control is lower. Conversely, at high immune recruitment rates and low tumor vascular functionality, the tumor can be controlled in a range of initial conditions, which is consistent with expectations. The initial concentration of effector cells $E$ only weakly depends on the immune recruitment rate $r$, but mostly on the level of functional tumor vascularization $B$, see Fig.~\ref{fig:figure9}(C,D). Concentrations of effector cells associated with the spiral and the saddle points both increase with increasing tumor vasculature, and approximately to the same extent, which corresponds to a parallel shift on the $E$-axis of both critical fixed points, see Fig.~\ref{fig:figure9}(C,D).

\subsection*{Bifurcation analysis}

In order to explore the critical parameter values at which the qualitative long-term dynamics of tumor evolution changes, we generated one-parameter bifurcation diagrams for the functional tumor vasculature $B$ and immune recruitment $r$, with the tumor radius plotted as a function of such parameters. Stable and unstable solution branches are represented by solid and dashed lines respectively. The Hopf and limit bifurcation points are indicated by $H$ and $LP$ respectively. A Hopf point is a local bifurcation in which a fixed point of a dynamic system loses stability as a pair of complex conjugate eigenvalues of the linearization around the fixed point cross the imaginary axis of the complex plane, i.e.~when the eigenvalues are a complex conjugate pair whose real part changes sign. The bifurcation analysis has been conducted by a prediction-correction continuation algorithm using the software MATCONT \cite{Dhooge2003}. 

\begin{figure}[H]
\centering
  \includegraphics[width=0.75\textwidth]{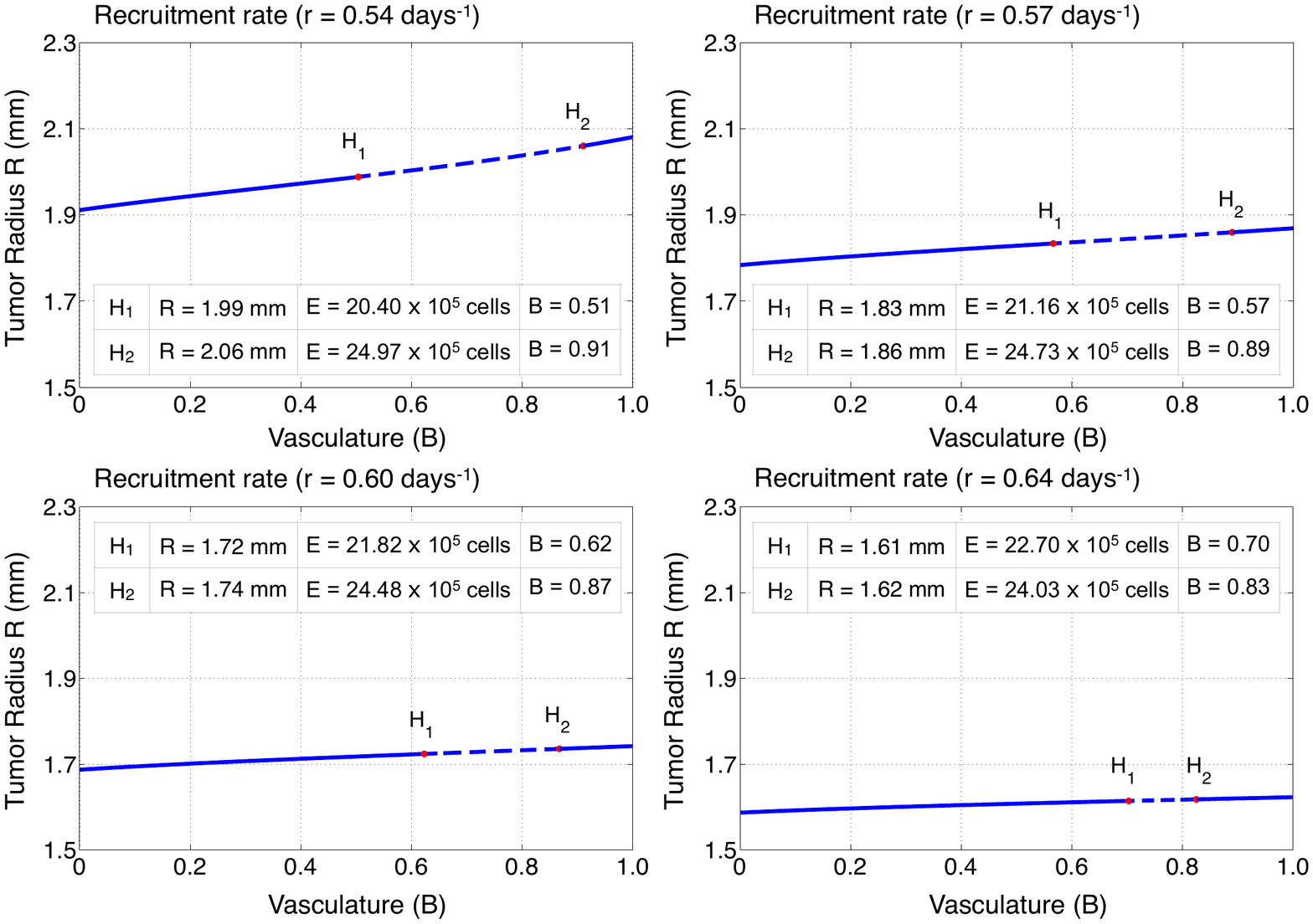}
  \caption{{\bf Bifurcation diagrams with respect to the functional tumor-associated vasculature $B$.} One-parameter bifurcation diagrams for different values of the immune recruitment rate $r = \{0.54,~0.57,~0.60,~0.64\}$~day$^{-1}$ in reading order. The labels $H_1$ and $H_2$ represent Hopf bifurcation points, while the solid and dashed lines describe the stable and unstable solution branches respectively. The concentration of effector cells $E$ corresponding to $H_1$ and $H_2$ are also provided. Remaining parameters are as in Tab.~1 of the manuscript.}
  \label{fig:figure10}
\end{figure}

For functional levels of the tumor-associated vasculature in a range around $B = 0.75$ and bound by the Hopf bifurcation points $H_1$ and $H_2$, the spiral fixed point is a repellor, see Fig.~\ref{fig:figure10}. Consequently, for $B$ in this range and for the different values of $r$, uncontrolled tumor growth occurs. Out of this critical range, i.e.~for lower and higher values of $B$, the spiral fixed point is an attractor, where long-term tumor control is reached. Not only low levels of functional vasculature allows for long-term tumor control, but there exists a therapeutic window of opportunity for well-vascularized tumors, see Fig.~\ref{fig:figure10}. With increasing immune recruitment rate $r$, the distance between the Hopf bifurcation points decreases, implying that the observed window of opportunity increases, while at the same time the global region of unfavourable situations decreases.

\begin{figure}[H]
\centering
  \includegraphics[width=0.75\textwidth]{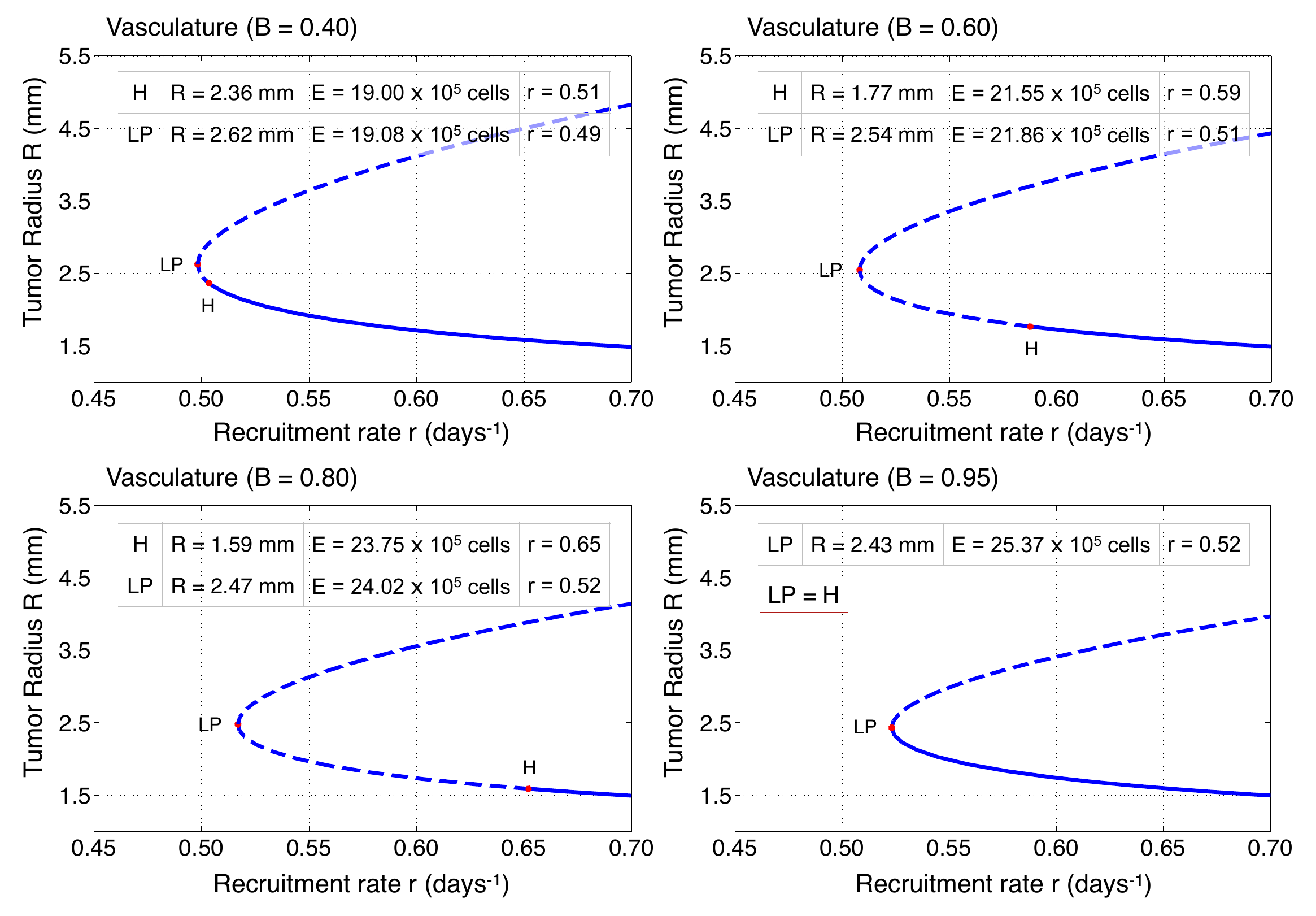}
  \caption{{\bf Bifurcation diagrams with respect to the immune recruitment rate $r$.} One-parameter bifurcation diagrams for different values of functional tumor-associated vasculature $B = \{0.40,~0.60,~0.80,~0.95\}$ in reading order. The labels $H$ and $LP$ represent a Hopf and a limit bifurcation point respectively. The solid and dashed lines describe the stable and unstable solution branches respectively. The concentration of effector cells $E$ corresponding to $H$ and $LP$ are also provided. Remaining parameters are as in Tab.~1 of the manuscript.}
  \label{fig:figure11}
\end{figure}

In the case of one-parameter bifurcation diagrams with respect to the immune recruitment rate $r$, see Fig.~\ref{fig:figure11}, we observe a limit and a Hopf bifurcation point. When tumor vascular functionality $B$ increases, the stable branch is gradually reduced and the Hopf bifurcation point moves from low to high values of $r$. However, there exists a value of $B$, at which this behaviour is inverted and the stable branch gets larger again. The range of $r$ values associated with the stable branch for $B = 0.95$ is similar to the one obtained for low values of $B$, for instance $B = 0.40$, see Fig.~\ref{fig:figure11}. In other words, even for a tumor with a highly functional vascular network, long-term control can be achieved for a wide range of immune recruitment rates. This finding suggests an alternative therapeutic strategy to the conventional antiangiogenic treatments. 

\subsection*{Estimation of the critical radius for tumor dormancy}

We derive an analytical expression that approximates the $R$-$E$ axis position of the saddle point, which corresponds to an upper bound estimation. $R \gg L_D$ and $B \rightarrow 1$ imply in Eq.~(2) in the manuscript that $\frac{L_D}{R} \rightarrow 0$ and $\tanh\left(R/L_{D} \right) \rightarrow 1$, as well as that $\frac{R^B}{R^{(B - 1)} + 1} \rightarrow \frac{R}{2}$. These simplifications and the quasi-steady state approximation of Eq.~(2) yields

\begin{equation*}\label{eq:ApproxE1}
    \frac{1}{3} (\lambda_M - \lambda_A) R - c E \frac{R}{2} = 0,
\end{equation*}

\noindent which implies that

\begin{equation*}\label{eq:ApproxE2}
    E = \frac{2 (\lambda_M - \lambda_A)}{3c}.
\end{equation*}

Arguing similarly for Eq.~(3) in the manuscript and under the same assumptions, we have that $\frac{R^3}{K + R^3} \rightarrow 1$ and $\frac{R^3}{1 + R^{(1 - B)}} \rightarrow \frac{R^3}{2}$, where

\begin{equation*}\label{eq:ApproxE3}
    r E - \left( d_1 \frac{R^3}{2} + d_0 \right) E + \sigma = 0,
\end{equation*}

\noindent and then we obtain that

\begin{equation*}\label{eq:ApproxE4}
    E = - \frac{\sigma}{r - \left( d_1 \frac{R^3}{2} + d_0 \right)}.
\end{equation*}

Combination of the previous equations for $E$ results in an approximation of the value of $R$ corresponding to the critical tumor radius, $R$-axis position of the saddle point, given by

\begin{equation*}\label{eq:ApproxR}
    R_S = \sqrt[3]{ \frac{3 \sigma c}{d_1 (\lambda_M - \lambda_A)} - \frac{2 (d_0 - r)}{d_1}}.
\end{equation*}


\subsection*{Estimation of control range of effector cell concentration}

We define the control range of effector cell concentration $\Delta E$ as the maximum difference of the $E$-axis spiral point position and the number of effectors $E^{*}$ given by the separatrix that defines sharp boundary between tumor dormant states and uncontrollable growth. Combinations of estimated $\Delta E$ with the position of the critical fixed points in the phase space for a given set of model parameters provide information about the ability of the immune responses to induce long-term tumor control. Values of initial concentration of effector cells $E_0$ out of $\Delta E$ will not result in long-term tumor dormancy, which suggests limitations of therapeutic immuno-modulatory interventions.



\begin{thebibliography}{10}

\bibitem{Folkman1972}
J.~Folkman, ``Anti-angiogenesis: new concept for therapy of solid tumors,''
  {\em Ann Surg}, vol.~175, pp.~409--416, Mar. 1972.

\bibitem{Jain2001}
R.~Jain, ``Normalizing tumor vasculature with anti-angiogenic therapy: A new
  paradigm for combination therapy,'' {\em Nat Med}, vol.~7, pp.~987--989,
  Sept. 2001.

\bibitem{Miao2012}
Z.~Miao, J.~Feng, and J.~Ding, ``Newly discovered angiogenesis inhibitors and
  their mechanisms of action,'' {\em Acta Pharmacol Sin}, vol.~33,
  pp.~1103--1111, Sept. 2012.

\bibitem{Argyriou2009}
A.~Argyriou, E.~Giannopoulou, and H.~Kalofonos, ``Angiogenesis and
  anti-angiogenic molecularly targeted therapies in malignant gliomas,'' {\em
  Oncology}, vol.~77, pp.~1--11, May 2009.

\bibitem{Ebos2011}
J.~Ebos and R.~Kerbel, ``Antiangiogenic therapy: impact on invasion, disease
  progression, and metastasis,'' {\em Nat Rev Clin Oncol}, vol.~8,
  pp.~210--221, Mar. 2011.

\bibitem{Jayson2012}
G.~Jayson, D.~Hicklin, and L.~Ellis, ``Antiangiogenic therapy--evolving view
  based on clinical trial results,'' {\em Nat Rev Clin Oncol}, vol.~9,
  pp.~297--303, Feb. 2012.

\bibitem{Jain2005}
R.~Jain, ``Normalization of tumor vasculature: An emerging concept in
  antiangiogenic therapy,'' {\em Science}, vol.~307, pp.~58--62, Jan. 2005.

\bibitem{Jain2013}
R.~Jain, ``Normalizing tumor microenvironment to treat cancer: bench to bedside
  to biomarkers,'' {\em J Clin Oncol}, vol.~31, pp.~2205--2218, June 2013.

\bibitem{Duda2007}
D.~Duda, R.~Jain, and C.~Willett, ``Antiangiogenics: the potential role of
  integrating this novel treatment modality with chemoradiation for solid
  cancers,'' {\em J Clin Oncol}, vol.~25, pp.~4033--4042, Sept. 2007.

\bibitem{Huang2012}
Y.~Huang, J.~Yuan, E.~Righi, W.~Kamoun, M.~Ancukiewicz, J.~Nezivar,
  M.~Santosuosso, J.~Martin, M.~Martin, F.~Vianello, P.~Leblanc, L.~Munn,
  P.~Huang, D.~Duda, D.~Fukumura, R.~Jain, and M.~Poznansky, ``Vascular
  normalizing doses of antiangiogenic treatment reprogram the immunosuppressive
  tumor microenvironment and enhance immunotherapy,'' {\em Proc Natl Acad Sci U
  S A}, vol.~109, pp.~17561--17566, Oct. 2012.

\bibitem{Huang2013}
Y.~Huang, S.~Goel, D.~Duda, D.~Fukumura, and R.~Jain, ``Vascular normalization
  as an emerging strategy to enhance cancer immunotherapy,'' {\em Cancer Res},
  vol.~73, pp.~2943--2948, May 2013.

\bibitem{Ferrara2005}
N.~Ferrara and R.~Kerbel, ``Angiogenesis as a therapeutic target,'' {\em
  Nature}, vol.~438, pp.~967--974, Dec. 2005.

\bibitem{Weis2011}
S.~Weis and D.~Cheresh, ``Tumor angiogenesis: molecular pathways and
  therapeutic targets,'' {\em Nat Med}, vol.~17, pp.~1359--1370, Nov. 2011.

\bibitem{Farnsworth2014}
R.~Farnsworth, M.~Lackmann, M.~Achen, and S.~Stacker, ``Vascular remodeling in
  cancer,'' {\em Oncogene}, vol.~33, pp.~3496--3505, Aug. 2014.

\bibitem{Fridman2012}
W.~Fridman, F.~Pag\`es, C.~Saut\`es-Fridman, and J.~Galon, ``The immune
  contexture in human tumours: impact on clinical outcome,'' {\em Nat Rev
  Cancer}, vol.~12, pp.~298--306, Mar. 2012.

\bibitem{Junttila2013}
M.~Junttila and F.~de~Sauvage, ``Influence of tumour micro-environment
  heterogeneity on therapeutic response,'' {\em Nature}, vol.~501,
  pp.~346--354, Sept. 2013.

\bibitem{Bellomo2008}
N.~Bellomo and M.~Delitala, ``From the mathematical kinetic, and stochastic
  game theory to modelling mutations, onset, progression and immune competition
  of cancer cells,'' {\em Phys Life Rev}, vol.~5, pp.~183--206, Dec. 2008.

\bibitem{Arciero2004}
J.~Arciero, T.~Jackson, and D.~Kirschner, ``A mathematical model of
  tumor-immune evasion and sirna treatment,'' {\em Discrete Continuous Dyn Syst
  Ser B}, vol.~4, pp.~39--58, Feb. 2004.

\bibitem{Matzavinos2004}
A.~Matzavinos, M.~Chaplain, and V.~Kuznetsov, ``Mathematical modelling of the
  spatio-temporal response of cytotoxic t-lymphocytes to a solid tumour,'' {\em
  Math Med Biol}, vol.~21, pp.~1--34, Mar. 2004.

\bibitem{DOnofrio2005}
A.~d'Onofrio, ``A general framework for modeling tumor-immune system
  competition and immunotherapy: Mathematical analysis and biomedical
  inferences,'' {\em Physica D: Nonlinear Phenomena}, vol.~208, pp.~220--235,
  Sept. 2005.

\bibitem{RobertsonTessi2012}
M.~Robertson-Tessi, A.~El-Kareh, and A.~Goriely, ``A mathematical model of
  tumor-immune interactions,'' {\em J Theor Biol}, vol.~294, pp.~56--73, Feb.
  2012.

\bibitem{Caravagna2010}
G.~Caravagna, A.~d'Onofrio, P.~Milazzo, and R.~Barbuti, ``Tumour suppression by
  immune system through stochastic oscillations,'' {\em J Theor Biol},
  vol.~265, pp.~336--345, Aug. 2010.

\bibitem{Koebel2007}
C.~Koebel, W.~Vermi, J.~Swann, N.~Zerafa, S.~Rodig, L.~Old, M.~Smyth, and
  R.~Schreiber, ``Adaptive immunity maintains occult cancer in an equilibrium
  state,'' {\em Nature}, vol.~450, pp.~903--907, 2007.

\bibitem{Pardoll2003}
D.~Pardoll, ``Does the immune system see tumors as foreign or self?,'' {\em
  Annu Rev Immunol}, vol.~21, pp.~807--39, 2003.

\bibitem{Dunn2004}
G.~Dunn, L.~Old, and R.~Schreiber, ``The three es of cancer immunoediting,''
  {\em Annu Rev Immunol}, vol.~22, pp.~329--60, 2004.

\bibitem{Stark2007}
J.~Stark, C.~Chan, and A.~George, ``Oscillations in the immune system,'' {\em
  Immunol Rev}, vol.~216, pp.~213--231, Apr. 2007.

\bibitem{Coventry2009}
B.~Coventry, M.~Ashdown, M.~Quinn, S.~Markovic, S.~Yatomi-Clarke, and
  A.~Robinson, ``Crp identifies homeostatic immune oscillations in cancer
  patients: a potential treatment targeting tool?,'' {\em J Transl Med},
  vol.~7, p.~102, 2009.

\bibitem{Kuznetsov1994}
V.~Kuznetsov, I.~Makalkin, M.~Taylor, and A.~Perelson, ``Nonlinear dynamics of
  immunogenic tumors: Parameter estimation and global bifurcation analysis,''
  {\em Bull Math Biol}, vol.~56, pp.~295--321, Mar. 1994.

\bibitem{Pillis2006}
L.~de~Pillis, W.~Gu, and A.~Radunskaya, ``Mixed immunotherapy and chemotherapy
  of tumors: modeling, applications and biological interpretations,'' {\em J
  Theor Biol}, vol.~238, pp.~841--862, Feb. 2006.

\bibitem{DOnofrio2010}
A.~d'Onofrio, F.~Gatti, P.~Cerrai, and L.~Freschi, ``Delay-induced oscillatory
  dynamics of tumourÐimmune system interaction,'' {\em Math Comput Model},
  vol.~51, pp.~572--591, Mar. 2010.

\bibitem{Araujo2004}
R.~Araujo and D.~McElwain, ``A history of the study of solid tumour growth: the
  contribution of mathematical modelling,'' {\em Bull Math Biol}, vol.~66,
  pp.~1039--1091, Sept. 2004.

\bibitem{Byrne2006}
H.~Byrne, T.~Alarcon, M.~Owen, S.~Webb, and P.~Maini, ``Modelling aspects of
  cancer dynamics: a review,'' {\em Philos Trans A Math Phys Eng Sci},
  vol.~364, pp.~1563--1578, June 2006.

\bibitem{Roose2007}
T.~Roose, S.~Chapman, and P.~Maini, ``Mathematical models of avascular tumor
  growth,'' {\em SIAM Rev}, vol.~49, pp.~179--208, May 2007.

\bibitem{Eftimie2011}
R.~Eftimie, J.~Bramson, and D.~Earn, ``Interactions between the immune system
  and cancer: a brief review of non-spatial mathematical models,'' {\em Bull
  Math Biol}, vol.~73, pp.~2--32, Jan. 2011.

\bibitem{Wilkie2013}
K.~Wilkie, ``A review of mathematical models of cancer-immune interactions in
  the context of tumor dormancy,'' {\em Adv Exp Med Biol}, vol.~734,
  pp.~201--234, Aug. 2013.

\bibitem{Schaller2005}
G.~Schaller and M.~Meyer-Hermann, ``Multicellular tumor spheroid in an
  off-lattice voronoi/delaunay cell model,'' {\em Phys Rev E}, vol.~71,
  pp.~051910--1--16, 2005.

\bibitem{Schaller2007}
G.~Schaller and M.~Meyer-Hermann, ``A modelling approach towards an epidermal
  homoeostasis control,'' {\em J Theor Biol}, vol.~247, pp.~554--573, Aug.
  2007.

\bibitem{Sachs2001}
R.~Sachs, L.~Hlatky, and P.~Hahnfeldt, ``Simple ode models of tumor growth and
  anti-angiogenic or radiation treatment,'' {\em Math Comput Model}, vol.~33,
  pp.~1297--1305, June 2001.

\bibitem{Byrne2010}
H.~Byrne, ``Dissecting cancer through mathematics: from the cell to the animal
  model,'' {\em Nat Rev Cancer}, vol.~10, pp.~221--230, Mar. 2010.

\bibitem{Jain2011}
H.~V. Jain and M.~Meyer-Hermann, ``The molecular basis of synergism between
  carboplatin and abt-737 therapy targeting ovarian carcinomas,'' {\em Cancer
  Res}, vol.~71, pp.~705--715, 2011.

\bibitem{Corwin2013}
D.~Corwin, C.~Holdsworth, R.~Rockne, A.~Trister, M.~Mrugala, J.~Rockhill,
  R.~Stewart, M.~Phillips, and K.~Swanson, ``Toward patient-specific,
  biologically optimized radiation therapy plans for the treatment of
  glioblastoma,'' {\em PLoS ONE}, vol.~8, p.~e79115, Nov. 2013.

\bibitem{Kempf2013}
H.~Kempf, H.~Hatzikirou, M.~Bleicher, and M.~Meyer-Hermann, ``In silico
  analysis of cell cycle synchronisation effects in radiotherapy of tumour
  spheroids,'' {\em PLoS Comput Biol}, vol.~9, p.~e1003295, Nov. 2013.

\bibitem{Alfonso2014a}
J.~Alfonso, N.~Jagiella, L.~N\'{u}\~{n}ez, M.~Herrero, and D.~Drasdo,
  ``Estimating dose painting effects in radiotherapy: a mathematical model,''
  {\em PLoS ONE}, vol.~9, p.~e89380, Feb. 2014.

\bibitem{Alfonso2014b}
J.~Alfonso, G.~Buttazzo, B.~Garc\'{i}a-Archilla, M.~Herrero, and
  L.~N\'{u}\~{n}ez, ``Selecting radiotherapy dose distributions by means of
  constrained optimization problems,'' {\em Bull Math Biol}, vol.~76,
  pp.~1017--1044, May 2014.

\bibitem{Jain2014}
H.~V. Jain, A.~Richardson, M.~Meyer-Hermann, and H.~M. Byrne, ``Exploiting the
  synergy between carboplatin and abt-737 in the treatment of ovarian
  carcinomas,'' {\em PLoS One}, vol.~9, p.~e81582, 2014.

\bibitem{Greenspan1976}
H.~Greenspan, ``On the growth and stability of cell cultures and solid
  tumors,'' {\em J Theor Biol}, vol.~56, pp.~229--242, Jan. 1976.

\bibitem{Byrne1995}
H.~Byrne and M.~Chaplain, ``Growth of nonnecrotic tumours in the presence and
  absence of inhibitors,'' {\em Math Biosci}, vol.~130, pp.~151--181, Dec.
  1995.

\bibitem{Cristini2003}
V.~Cristini, J.~Lowengrub, and Q.~Nie, ``Nonlinear simulation of tumor
  growth,'' {\em J Math Biol}, vol.~46, pp.~191--224, Mar. 2003.

\bibitem{Hatzikirou2012}
H.~Hatzikirou, A.~Chauvi\`{e}re, J.~Lowengrub, J.~De~Groot, and V.~Cristini,
  ``Effect of vascularization on glioma tumor growth,'' in {\em Modeling Tumor
  Vasculature} (T.~L. Jackson, ed.), pp.~237--259, Springer New York, 2012.

\bibitem{DePillis2005}
L.~de~Pillis, A.~Radunskaya, and C.~Wiseman, ``A validated mathematical model
  of cell-mediated immune response to tumor growth,'' {\em Cancer Res},
  vol.~65, pp.~7950--8, Sept. 2005.

\bibitem{Herman2011}
A.~Herman, V.~Savage, and G.~West, ``{A quantitative theory of solid tumor
  growth, metabolic rate and vascularization},'' {\em PLoS ONE}, vol.~6, no.~9,
  p.~e22973, 2011.

\bibitem{Stauffer1994}
D.~Stauffer and A.~Aharony, {\em Introduction To Percolation Theory}.
\newblock Taylor \& Francis, 1994.

\bibitem{Paglia1996}
P.~Paglia, C.~Chiodoni, M.~Rodolfo, and M.~Colombo, ``Murine dendritic cells
  loaded in vitro with soluble protein prime cytotoxic t lymphocytes against
  tumor antigen in vivo.,'' {\em J Exp Med}, vol.~183, no.~1, pp.~317--322,
  1996.

\bibitem{Bru2004}
A.~Br\'{u}, S.~Albertos, J.~Garc\'{\i}a-Asenjo, and I.~Br\'{u}, ``Pinning of
  tumoral growth by enhancement of the immune response,'' {\em Phys Rev Lett},
  vol.~92, pp.~1--4, June 2004.

\bibitem{Davison1997}
A.~Davison and D.~Hinkley, {\em Bootstrap Methods and their Application}.
\newblock Cambridge: Cambridge University Press, 1997.

\bibitem{Alessandri2013}
K.~Alessandri, B.~Sarangi, V.~Gurchenkov, B.~Sinha, T.~Kie$\ss$ling, L.~Fetler,
  F.~Rico, S.~Scheuring, C.~Lamaze, A.~Simon, S.~Geraldo, D.~Vignjevic,
  H.~Dom\'ejean, L.~Rolland, A.~Funfak, J.~Bibette, N.~Bremond, and P.~Nassoy,
  ``Cellular capsules as a tool for multicellular spheroid production and for
  investigating the mechanics of tumor progression in vitro,'' {\em Proc Natl
  Acad Sci U S A}, vol.~110, pp.~14843--14848, Aug. 2013.

\bibitem{Delarue2013}
M.~Delarue, F.~Montel, O.~Caen, J.~Elgeti, J.~Siaugue, D.~Vignjevic, J.~Prost,
  J.~Joanny, and G.~Cappello, ``Mechanical control of cell flow in
  multicellular spheroids,'' {\em Phys Rev Lett}, vol.~110, p.~138103, Mar.
  2013.

\bibitem{Frieboes2006}
H.~Frieboes, X.~Zheng, C.~Sun, B.~Tromberg, R.~Gatenby, and V.~Cristini, ``An
  integrated computational/experimental model of tumor invasion,'' {\em Cancer
  Res}, vol.~66, pp.~1597--1604, Feb. 2006.

\bibitem{Cristini2008}
V.~Cristini, H.~Frieboes, X.~Li, J.~Lowengrub, P.~Macklin, S.~Sanga, S.~Wise,
  and X.~Zheng, ``Nonlinear modeling and simulation of tumor growth,'' in {\em
  Selected topics in cancer modeling: Genesis, evolution, immune competition,
  and therapy. Modelling and Simulation in Science, Engineering, and
  Technology} (N.~Bellomo, M.~Chaplain, and E.~de~Angelis, eds.), ch.~6,
  pp.~113--82, Boston, MA USA: Birkh{\"a}user, 2008.

\bibitem{Su2009}
B.~Su, W.~Zhou, K.~Dorman, and D.~Jones, ``Mathematical modelling of immune
  response in tissues,'' {\em Comput Math Methods Med}, vol.~10, no.~1,
  pp.~9--38, 2009.

\bibitem{DOnofrio2012}
A.~d'Onofrio, U.~Ledzewicz, and H.~Sch{\"a}ttler, ``On the dynamics of
  tumor-immune system interactions and combined chemo-and immunotherapy,'' in
  {\em New Challenges for Cancer Systems Biomedicine}, pp.~249--266, Springer,
  2012.

\bibitem{Seshadri2007}
M.~Seshadri, J.~Spernyak, P.~Maiery, R.~Cheney, R.~Mazurchuk, and D.~Bellnier,
  ``Visualizing the acute effects of vascular-targeted therapy in vivo using
  intravital microscopy and magnetic resonance imaging: correlation with
  endothelial apoptosis, cytokine induction, and treatment outcome,'' {\em
  Neoplasia}, vol.~9, pp.~128--35, Feb. 2007.

\bibitem{Nocedal2006}
J.~Nocedal and S.~Wright in {\em Numerical Optimization}, Springer Series in
  Operations Research and Financial Engineering, Springer-Verlag New York,
  2006.

\bibitem{Hyun2013}
A.~Hyun and P.~Macklin, ``Improved patient-specific calibration for agent-based
  cancer modeling,'' {\em J Theor Biol}, no.~317, pp.~422--424, 2013.

\bibitem{Hoeben2004}
A.~Hoeben, B.~Landuyt, M.~Highley, H.~Wildiers, A.~Van~Oosterom, and
  E.~De~Bruijn, ``Vascular endothelial growth factor and angiogenesis,'' {\em
  Pharmacological Reviews}, vol.~56, pp.~549--580, Dec. 2004.

\bibitem{Quail2013}
D.~Quail and J.~Joyce, ``Microenvironmental regulation of tumor progression and
  metastasis,'' {\em Nat Med}, vol.~19, pp.~1423--1437, Nov. 2013.

\bibitem{Goel2011}
S.~Goel, D.~Duda, L.~Xu, L.~Munn, Y.~Boucher, D.~Fukumura, and R.~Jain,
  ``Normalization of the vasculature for treatment of cancer and other
  diseases,'' {\em Physiol Rev}, vol.~91, pp.~1071--1121, July 2011.

\bibitem{Stylianopoulos2012}
T.~Stylianopoulos, J.~Martin, V.~Chauhan, S.~Jain, B.~Diop-Frimpong,
  N.~Bardeesy, B.~Smith, C.~Ferrone, F.~Hornicek, Y.~Boucher, L.~Munn, and
  J.~{R. K.}, ``Causes, consequences, and remedies for growth-induced solid
  stress in murine and human tumors,'' {\em Proc Natl Acad Sci U S A},
  vol.~109, pp.~15101--15108, Aug. 2012.

\bibitem{Manning2007}
E.~Manning, J.~Ullman, J.~Leatherman, J.~Asquith, T.~Hansen, T.~Armstrong,
  D.~Hicklin, E.~Jaffee, and L.~Emens, ``A vascular endothelial growth factor
  receptor-2 inhibitor enhances antitumor immunity through an immune-based
  mechanism,'' {\em Clin Cancer Res}, vol.~13, pp.~3951--3959, July 2007.

\bibitem{Hamzah2008}
J.~Hamzah, M.~Jugold, F.~Kiessling, P.~Rigby, M.~Manzur, H.~Marti, T.~Rabie,
  S.~Kaden, H.~Gr{\"o}ne, G.~H{\"a}mmerling, B.~Arnold, and R.~Ganss,
  ``Vascular normalization in rgs5-deficient tumours promotes immune
  destruction,'' {\em Nature}, vol.~453, pp.~410--414, May 2008.

\bibitem{Shrimali2010}
R.~Shrimali, Z.~Yu, M.~Theoret, D.~Chinnasamy, N.~Restifo, and S.~Rosenberg,
  ``Antiangiogenic agents can increase lymphocyte infiltration into tumor and
  enhance the effectiveness of adoptive immunotherapy of cancer,'' {\em Cancer
  Res}, vol.~70, pp.~6171--6180, Aug. 2010.

\bibitem{Agrawal2014}
V.~Agrawal, S.~Maharjan, K.~Kim, N.~Kim, J.~Son, K.~Lee, H.~Choi, S.~Rho,
  S.~Ahn, M.~Won, S.~Ha, G.~Koh, Y.~Kim, Y.~Suh, and Y.~Kwon, ``Direct
  endothelial junction restoration results in significant tumor vascular
  normalization and metastasis inhibition in mice,'' {\em Oncotarget}, vol.~5,
  no.~9, pp.~2761--77, 2014.

\bibitem{Macklin2007}
P.~Macklin and J.~Lowengrub, ``Nonlinear simulation of the effect of
  microenvironment on tumor growth,'' {\em J Theor Biol}, vol.~245, no.~4,
  pp.~677--704, 2007.

\bibitem{Visser2006}
K.~de~Visser, A.~Eichten, and L.~Coussens, ``Paradoxical roles of the immune
  system during cancer development,'' {\em Nat Rev Cancer}, vol.~6, pp.~24--37,
  Jan. 2006.

\bibitem{Hao2012}
N.~Hao, M.~L{\"u}, Y.~Fan, Y.~Cao, Z.~Zhang, and S.~Yang, ``Macrophages in
  tumor microenvironments and the progression of tumors,'' {\em Clin Dev
  Immunol}, vol.~2012, p.~11, June 2012.

\bibitem{Acker1984}
H.~Acker, {\em Spheroids in cancer research: methods and perspectives}.
\newblock Springer-Verlag Berlin; New York, 1984.

\bibitem{Dhooge2003}
A.~Dhooge, W.~Govaerts, and Y.~A. Kuznetsov, ``Matcont: A matlab package for
  numerical bifurcation analysis of odes,'' {\em ACM Trans Math Softw},
  vol.~29, pp.~141--164, June 2003.

\end{thebibliography}

\bibliographystyle{plain}

\end{document}